\documentclass[aps,floats,floatfix,showpacs,amssymb,prd,twocolumn,superscriptaddress,nofootinbib,nolongbibliography,10pt]{revtex4-2}

\usepackage{amssymb,amsmath,verbatim,mathtools,needspace,enumitem,etoolbox,graphicx,physics,microtype,afterpage,bigints,textcomp,gensymb,tabularx,bm,booktabs,multirow,ragged2e,nicefrac,colortbl,scalerel,comment}

\usepackage[dvipsnames]{xcolor}
\definecolor{linkcolor}{rgb}{0.0,0.3,0.5}
\definecolor{lightblue}{RGB}{200,220,240}
\definecolor{dodgerblue}{HTML}{1E90FF}
\definecolor{MidnightBlueLight}{RGB}{239,239,251}
\usepackage[unicode, colorlinks=true, linkcolor=linkcolor, citecolor=linkcolor, filecolor=linkcolor,urlcolor=linkcolor, pdfusetitle]{hyperref}
\usepackage[all]{hypcap}
\usepackage[T1]{fontenc}
\usepackage[utf8]{inputenc}
\usepackage{orcidlink}
\makeatletter
\newcommand{\partialvardoiint}{\mathop{\mathpalette\make@partialvardoiint\relax}\nolimits}
\newcommand{\make@partialvardoiint}[2]{%
  \begin{tikzpicture}[baseline=(int.base), inner sep=0, outer sep=0]    
    \node (int) at (0,0) {};
    \def\stadiumwidth{0.55em}
    \def\stadiumheight{0.45em}
    \def\stadiumrounded{0.24em}
    
    \ifx#1\displaystyle
      \def\sc{1.0}
    \else\ifx#1\textstyle
      \def\sc{0.9}
    \else
      \def\sc{0.7}
    \fi\fi
    \draw[line width=0.04em, rounded corners=\stadiumrounded] 
      ([shift={(-\stadiumwidth/2, -\stadiumheight/2)}]int.center) 
      rectangle 
      ([shift={(\stadiumwidth/2, \stadiumheight/2)}]int.center);
    
  \end{tikzpicture}
}
\makeatother
\makeatletter
\newcommand*{\balancecolsandclearpage}{\close@column@grid \cleardoublepage \twocolumngrid}
\makeatother

\usepackage{lipsum}

\newcommand{\jhu}{\affiliation{William H. Miller III Department of Physics and Astronomy,\\ Johns Hopkins University, 3400 North Charles Street, Baltimore, Maryland, 21218, USA}}

\newcommand{\UTAustin}{\affiliation{Weinberg Institute, University of Texas at Austin, Austin, TX 78712, USA}}

\newcommand{\PennStateIGC}{\affiliation{Institute for Gravitation and the Cosmos, Department of Physics, Penn State University, University Park, Pennsylvania, 16801, USA}}

\newcommand{\PennStateDAA}{\affiliation{Department of Astronomy and Astrophysics, Penn State University, University Park, Pennsylvania, 16801, USA}}

\begin{document}

\title{Not too close! Evaluating the impact of the baseline on the localization\texorpdfstring{\\}{ }of binary black holes by next-generation gravitational-wave detectors}

\author{Francesco Iacovelli\texorpdfstring{\,}{ }\orcidlink{0000-0002-4875-5862}}\email{fiacovelli@jhu.edu}
\jhu

\author{Luca Reali\texorpdfstring{\,}{ }\orcidlink{0000-0002-8143-6767}}
\jhu

\author{Emanuele Berti\texorpdfstring{\,}{ }\orcidlink{0000-0003-0751-5130}}
\jhu

\author{Alessandra Corsi\texorpdfstring{\,}{ }\orcidlink{0000-0001-8104-3536}}
\jhu

\author{B. S. Sathyaprakash\texorpdfstring{\,}{ }\orcidlink{0000-0003-3845-7586}}
\PennStateIGC
\PennStateDAA

\author{Digvijay Wadekar\texorpdfstring{\,}{ }\orcidlink{0000-0002-2544-7533}}
\jhu
\UTAustin


\date{\today}

\begin{abstract}
	Next-generation (XG) gravitational-wave detectors, such as Cosmic Explorer (CE) and the Einstein Telescope (ET), will observe compact binary coalescences at unprecedented rates and signal-to-noise ratios. Accurate sky localization of these sources is crucial for several aspects of the science case of CE and ET. The localization of most binary black hole (BBH) signals, which will spend at most a few minutes within the XG detector’s effective sensitivity band, will continue to rely primarily on timing triangulation across a network of detectors. A key design choice for triangulation is the baseline between instruments. We investigate how the baseline affects the localization capabilities of a two-detector CE network, analyzing both fixed-parameter injections and a realistic BBH population consistent with the latest GWTC-4 results. For detector-frame total masses up to $\sim\!100\,{\rm M}_\odot$, we find that baselines corresponding to light travel times of $8-11$\,ms ($\sim\!2300-3300$\,km) offer a reasonable compromise, producing predominantly unimodal or bimodal sky localizations suitable for electromagnetic follow-up and statistical host galaxy identification and galaxy cross-correlation studies. Shorter baselines significantly degrade localization, particularly for high signal-to-noise ratio events. Crucially, we find that adding a third detector to the network eliminates localization multimodality for a substantial fraction of sources. A network with two CEs and LIGO-India provides unimodal posteriors for a good fraction of the events, while two CEs plus ET would provide unimodal posteriors for essentially all of them. These considerations should be useful to inform the development of the XG detector network.
\end{abstract}

\maketitle

\section{Introduction}

In the past decade, the observation of the gravitational-wave (GW) signal emitted by compact binary coalescences (CBCs) gave us a new tool to study the Universe. With more than 200 event candidates observed by the LIGO~\cite{LIGOScientific:2014pky}, Virgo~\cite{Virgo:2014yos}, and KAGRA~\cite{Aso:2013eba} (LVK) detectors, the GWs give us insight into the demography of masses, redshifts, and spins of compact object binaries~\cite{KAGRA:2021duu, LIGOScientific:2025pvj}, the nature of the objects that produce them~\cite{LIGOScientific:2021sio,LIGOScientific:2025rid}, their composition~\cite{LIGOScientific:2017vwq}, and the Universe through which they propagate~\cite{LIGOScientific:2021aug,LIGOScientific:2025jau}. 

The GW community is now preparing for the next step: next-generation (XG) ground-based detectors.
With a considerable increase in sensitivity and a broader frequency range compared to current instruments, the Einstein Telescope (ET) in Europe~\cite{Punturo:2010zz,Hild:2010id,ET:2025xjr} and Cosmic Explorer (CE) in the US~\cite{Reitze:2019iox,Evans:2021gyd,Evans:2023euw} have the potential to revolutionize astrophysics, fundamental physics, and cosmology. They are expected to observe most of the binary black holes (BBHs) and a large fraction of the binary neutron stars (BNSs) merging in the Universe out to very high redshifts, with detection rates as high as $\gtrsim10^5$ events per year~\cite{Borhanian:2022czq,Ronchini:2022gwk,Iacovelli:2022bbs,Pieroni:2022bbh,Gupta:2023lga}. 

A crucial aspect of the science case is the ability to accurately reconstruct the sky position and distance of the observed sources. Good sky localization would allow us to search for potential EM counterparts and facilitate host galaxy associations, obtain stringent constraints on the cosmic expansion history of the Universe through the so-called dark sirens method~\cite{Schutz:1986gp,MacLeod:2007jd,DelPozzo:2011vcw,LIGOScientific:2018gmd,Gray:2019ksv,LIGOScientific:2019zcs,Borhanian:2020vyr,Finke:2021aom,LIGOScientific:2021aug,Borghi:2023opd,LIGOScientific:2025jau}, reconstruct the evolution with redshift of the merger rate of compact binary systems~\cite{VanDenBroeck:2013uza,vanSon:2021zpk,Ng:2022agi,Branchesi:2023mws}, 
and identify binaries produced by high-redshift formation channels, such as Population III stars and primordial black holes (BHs)~\cite{Carr:1975qj,Kinugawa:2014zha,Hartwig:2016nde,Belczynski:2016ieo,Sasaki:2018dmp,Carr:2020gox,Franciolini:2021xbq,Mancarella:2023ehn,Franciolini:2023opt,Santoliquido:2024oqs,Mestichelli:2024djn,Franciolini:2024vis,Plunkett:2025mjr}. Binaries containing neutron stars (NSs), thanks to the baryonic material they expel, can produce EM counterparts across the spectrum, from $\gamma$-rays to radio, and on different time scales, from seconds to several months (see e.g. Refs.~\cite{Metzger:2019zeh,Nakar:2019fza,Ascenzi:2020xqi} and references therein), as confirmed by the multimessenger observation of GW170817~\cite{LIGOScientific:2017vwq,LIGOScientific:2017zic}. Mergers involving two BHs are instead generally expected to be electromagnetically ``dark'' in the absence of surrounding matter. Nevertheless, following the possible association of EM counterparts to the BBH mergers GW150914~\cite{Connaughton:2016umz} and GW190521~\cite{Graham:2020gwr}, several mechanisms resulting in an EM emission accompanying these events have been proposed, mostly based on accretion or shocking of surrounding material (e.g., for binaries in active galactic nuclei disks or with mini-disks)~\cite{Bartos:2016dgn,Stone:2016wzz,Perna:2016jqh,Ford:2019nic,McKernan:2019hqs,Veres:2019hsd,Graham:2020gwr,Kelly:2020vpv,Kimura:2021xxu}, or a charge of the BHs~\cite{Zhang:2016rli,Liebling:2016orx}. These can power $\gamma$-ray burst (GRB) emission on timescales of seconds to optical, UV, X-ray, and counterparts on timescales of days to months.

While XG detections will have higher signal-to-noise ratios (SNRs), allowing for an exquisite reconstruction of their intrinsic parameters, the determination of the extrinsic parameters (in particular, the sky location) does not depend only on the noise floor, but also on the number of available detectors, their baseline, and their relative orientation. Indeed, the localization of a signal observed by two detectors is known to be poor and potentially multimodal~\cite{Fairhurst:2009tc,Wen:2010cr,Fairhurst:2010is}: for example, the localization regions reconstructed by current interferometers when only two detectors are active are much broader~\cite{LIGOScientific:2018mvr,LIGOScientific:2020ibl,KAGRA:2021vkt,LIGOScientific:2025slb}.

The dominant contribution to the localization of a signal comes from the time-delay between different instruments, which allows triangulation of their sky position. For lighter binaries observed with XG detectors, an important improvement in sky localization will come from the duration of the signal in the detectors' sensitivity band. While GW170817~\cite{LIGOScientific:2017vwq}, the longest GW signal observed to date,  spent $\sim\!30\,{\rm s}$ within the LIGO-Virgo frequency range, an event with the same properties would be observable by ET for about a day~\cite{Iacovelli:2022bbs}. With such a long observation time, the motion of the detector due to Earth's rotation makes it possible to ``triangulate'' the signal even with a single detector~\cite{Nitz:2021pbr,Baral:2023xst,Baral:2025geo}. For short signals, and in particular for most BBHs, this will not be possible, potentially resulting in multimodal localization posteriors~\cite{Singh:2020wsy,Singh:2021bwn,Santoliquido:2025lot,Santoliquido:2025aiq}. 
In particular, the authors of Ref.~\cite{Santoliquido:2025aiq} showed that, for high mass and high redshift sources observed by ET in a configuration consisting of 2 L-shaped detectors located in Europe~\cite{Branchesi:2023mws}, the luminosity distance posterior might exhibit a multimodal structure due to correlations with a multimodal localization posterior. For such signals, one of the key aspects will be the available baseline between the detectors in a given network. A larger separation between the detectors results in a better reconstruction of the time delays between different instruments, thus improving the triangulation and mitigating the consequent multimodality on the source distance.  

In this paper, we analyze how the baseline affects the 3-d localization of BBH signals observed by two CE interferometers. We show that a moderate baseline of $\sim2300\,{\rm km}$ between two CE detectors operating in isolation is a reasonable compromise. Adding a third detector to the network will get rid of multimodal localization posteriors for a significant fraction of the source population detected by all three instruments. A network with two CEs and a LIGO-India observatory~\cite{LIGOIndia} can eliminate multimodal localization posteriors for many events up to detector frame total masses $M_{\rm tot}\lesssim200\,{\rm M}_\odot$. The best scenario consists of an XG network with two CEs and ET, as this can provide unimodal posteriors for essentially all of the events. 

The rest of the paper is organized as follows. In Sec.~\ref{sec:setup}, we describe detector networks, the forecasting tools, and our simulation setup. In Sec.~\ref{sec:results}, we present our results for various CE baselines and explore their dependence on the mass ratio and inclination of simulated binaries. We also present and discuss results for a realistic population of short BBH signals consistent with the latest GWTC-4 event catalog~\cite{LIGOScientific:2025pvj}. Finally, we discuss results for a global network including either a LIGO-India detector or ET. Sec.~\ref{sec:conclusions} contains our conclusions. Appendix~\ref{app:et} provides localization forecasts as a function of the baseline for two L-shaped ET detectors using the same methods, while Appendix~\ref{app:highermodes} shows the impact of including higher-order modes in a part of the analysis.

\section{Analysis setup}\label{sec:setup}

\subsection{Detector networks and sensitivity}\label{subsec:detectors}

We consider a pair of L-shaped detectors with the characteristics of CE. In particular, following Ref.~\cite{Evans:2023euw}, we consider a first detector with 40~km arm length and a second with 20~km arm length. We adopt the latest sensitivity curves publicly available in the \href{https://github.com/koustavchandra/gwforge}{\texttt{gwforge}} package~\cite{Chandra:2024dhf} in the most optimistic scenario, which corresponds to a high laser power of $1.5\,{\rm MW}$ and improved coatings~\cite{noise_curves_gwforge}. %
We consider the frequency range $f\in[5,\,2048]\,{\rm Hz}$, which is appropriate for BBH systems.

The crucial aspect for our study is the location of the two detectors. As a reference, we place the detectors at two extremal points on the US mainland separated by a great circle distance of approximately 4465~km, with a light travel time between the two of $\sim\!15\,{\rm ms}$. The 40~km detector is fixed at the north-west end of this curve, with the $x$-arm aligned with local East and the $y$-arm aligned with local North. %
We then vary the baseline between the two CEs by placing the 20~km detector at different points along the great circle connecting the two extremal locations. The intermediate points correspond to different representative distances (light travel times), namely: 595~km ($2\,{\rm ms}$), 1191~km ($4\,{\rm ms}$), 1786~km ($6\,{\rm ms}$), 2382~km ($8\,{\rm ms}$), and 3275~km ($11\,{\rm ms}$). The latter baseline is comparable to the case of the two LIGO detectors, which are separated by a great circle distance of $\sim\!3000\,{\rm km}$ and a light travel time of 10\,ms.
At each location, the 20~km CE is oriented with a relative angle of $45^\circ$ with respect to the great circle connecting it to the fixed 40~km CE, ensuring an optimal performance in terms of localization. Notice that the choice of misaligned detectors would in turn result in a significant loss in sensitivity to stochastic GW backgrounds (SGWBs)~\cite{Romano:2016dpx}. Given this is not the focus of our work, we will use the $45^\circ$, but the impact of the alignment on SGWB searches should be taken into account for the final design of the detector.    

\begin{figure}[tbp]
    \centering
    \includegraphics[width=\linewidth]{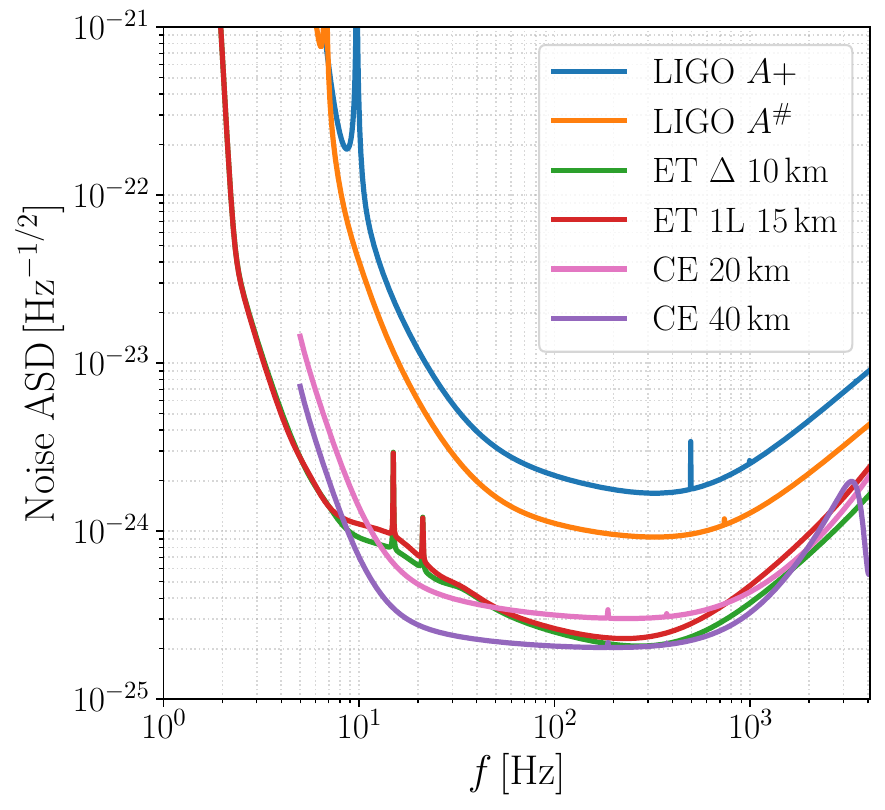}\\
    \caption{Noise amplitude spectral densities (ASDs) for the various detectors considered in this work. For the triangular ET configuration, we show the noise curve rescaled by a factor $2/3$ as a proxy of the detector geometry, see the text.}
    \label{fig:all_asds}
\end{figure}

In the second part of the analysis, we consider a global network consisting of the 2 CE detectors with the different chosen baselines, as well as one of the following detectors:
\begin{itemize}
    \item LIGO-India (LI) with sensitivity corresponding to either the $A+$ target sensitivity~\cite{KAGRA:2013rdx} or the $A^\#$ design~\cite{AsharpReport};
    \item ET in its triangular ($\Delta$) configuration, consisting of three nested interferometers with $60^\circ$ opening angle, located in Sardinia, Italy;
    \item ET in its 2L configuration with misaligned detectors, one located in Sardinia, Italy, and one in the Meuse-Rhine Euroregion~\cite{Branchesi:2023mws}. The great circle distance between the two detectors in this configuration is $\sim\!1167\,{\rm km}$, corresponding to a light travel time of $3.9\,{\rm ms}$.
\end{itemize}
The sensitivity curves used for LIGO-India can be found in Refs.~\cite{noise_curve_Aplusdes,noise_curve_Asharp}, while the location and sensitivity curves for ET follow Ref.~\cite{Branchesi:2023mws}. As in Ref.~\cite{Branchesi:2023mws}, we consider an arm length of 10\,km for the triangular ET geometry and 15\,km for the 2L ET geometry. When simulating events observed with ET, we adopt the frequency range $f\in[3,\,2048]\,{\rm Hz}$. The noise curves for each detector we consider are shown in Fig.~\ref{fig:all_asds}. For the triangular ET configuration, we show the noise curve rescaled by a factor $2/3$, which approximately takes into account the detector geometry (see e.g.~\cite{Iacovelli:2022bbs}).
In all the analyses, we assume a 100\% duty cycle for the detectors.

\subsection{Evaluation of the 3D sky localization and injections}\label{subsec:loc_method_injs}

To evaluate the sky localization capabilities of each pair of CE detectors, we employ the publicly available \href{https://git.ligo.org/lscsoft/ligo.skymap}{\texttt{ligo.skymap}} package~\cite{Singer:2015ema,Singer:2016eax,Singer:2016erz}, in particular the \href{https://git.ligo.org/lscsoft/ligo.skymap/-/tree/main/ligo/skymap/bayestar}{\texttt{BAYESTAR}} module. This allows us to go beyond the Fisher matrix approximation for the localization, which is widely used in the XG forecasting literature (see e.g. Refs.~\cite{Borhanian:2022czq,Ronchini:2022gwk,Pieroni:2022bbh,Iacovelli:2022bbs,Gupta:2023lga,Begnoni:2025oyd}), and study multimodalities in the sky position and luminosity distance posterior, which the Fisher approach cannot capture by construction. We have included our chosen detectors and noise curves in the LIGO Algorithm Library, \href{https://git.ligo.org/lscsoft/lalsuite}{\texttt{LAL}}~\cite{lalsuite}, making them accessible through \texttt{ligo.skymap}. 

We evaluate the sky-localization capabilities of each configuration as follows:
    
\begin{enumerate}
    \item Through the \href{https://git.ligo.org/lscsoft/lalsuite/-/tree/master/lalapps}{\texttt{LALApps}} \texttt{inspinj} module we generate injection files to be analyzed with \texttt{BAYESTAR}. We consider BBHs with 40 different detector-frame masses logarithmically distributed from $M_{\rm tot}^{\rm det}=20\,{\rm M}_\odot$ to $M_{\rm tot}^{\rm det}=1000\,{\rm M}_\odot$. In our default settings, we consider equal-mass systems (for which the symmetric mass ratio $\eta=1/4)$, and study the effect of the mass ratio on our analysis in Sec.~\ref{subsec:mass_ratio_var}. The chosen mass range ensures that the longest time spent in band for a signal that starts at frequency $f_l$ is~\cite{Sathyaprakash:1994nj}, 
    \begin{equation}\label{eq:time_to_coal}
        \tau(f_l)\approx \dfrac{5}{256\eta} (\pi f_l)^{-8/3}\left(\dfrac{GM_{\rm tot}}{c^3}\right)^{-5/3}\,,
    \end{equation}
    is at most $\tau(5\,{\rm Hz})\simeq4\,{\rm min}$, for which the impact of Earth's motion, not included in \texttt{BAYESTAR}, is negligible. When considering ET, given its lower frequency reach, we consider systems with masses in the interval $M_{\rm tot}\in[40,\,2000]\,{\rm M}_\odot$, such that the observation time for a signal is at most $\tau(3\,{\rm Hz})\simeq5\,{\rm min}$. In all the analyses, we consider nonspinning systems. We further neglect in the analysis the impact of a frequency dependence in the response of the detectors, restricting ourselves to the long-wavelength approximation. However, we note that for interferometers with arms as long as ET and CE, this could give a relevant contribution for signals merging already at a fraction of the free spectral range frequency, $f_{\rm fsr}= c/(2L_{\rm arm})$ with $L_{\rm arm}$ denoting the arm length of the instrument, and thus impact localization~\cite{Essick:2017wyl,Virtuoso:2024kyp}.
    \item For each value of the mass, we distribute the binaries uniformly in the sky on a $40\times40$ grid in right ascension and declination. 
    Regarding the other extrinsic parameters, in all the analyses we consider a fixed observation time $t_{\rm GPS} = 10^{9}\,{\rm s}$, vanishing polarization angle and phase, and in the default settings we consider face-on systems. We study the effect of a nonvanishing inclination in Sec.~\ref{subsec:inclination_var}.
    \item\label{item:snr_distr} For each system, we compute the optimal matched-filter SNR using the \href{https://github.com/CosmoStatGW/gwfast}{\texttt{gwfast}} package~\cite{Iacovelli:2022bbs,Iacovelli:2022mbg} at a pivot luminosity distance of $100\,{\rm Mpc}$. We then scale the distance to obtain a network SNR of 30, 60, or 120. These values are approximately the SNR values for the median, and upper 68\% and 95\% percentiles of the SNR distribution for a realistic 1\,yr BBH population observed by two CE detectors with the chosen sensitivity curves and the longest baseline considered. The SNR distribution is shown in Fig.~\ref{fig:snr_distr_CE} and obtained as follows. We extract 100 hyperparameter samples from the latest LVK population results~\cite{LIGOScientific:2025pvj}, adopting the \textsc{Broken Power Law + 2 Peaks} model for the source-frame component masses, \textsc{Gaussian Component Spins} for the spin magnitudes and tilts, and \textsc{Power Law Redshift} for the redshift, extending it up to $z=10$ using a Madau-Dickinson profile~\cite{Madau:2014bja} in which the low redshift slope is fixed at the value found by LVK, while the peak redshift and high-$z$ slope are set to the typical values $z_p=2$ and $\beta_z=3$~\cite{Madau:2016jbv}. The other parameters are sampled from uniform physical priors. For each hyperparameter sample, we generate a BBH catalog for 1\,yr of observation, and compute the SNR of each event in the catalog using \texttt{gwfast}. From each of these catalogs, we compute the median and upper 68\% and 95\% percentiles, and take the average values over the different realizations. 
    
    \begin{figure}[tb]
        \centering
        \includegraphics[width=\linewidth]{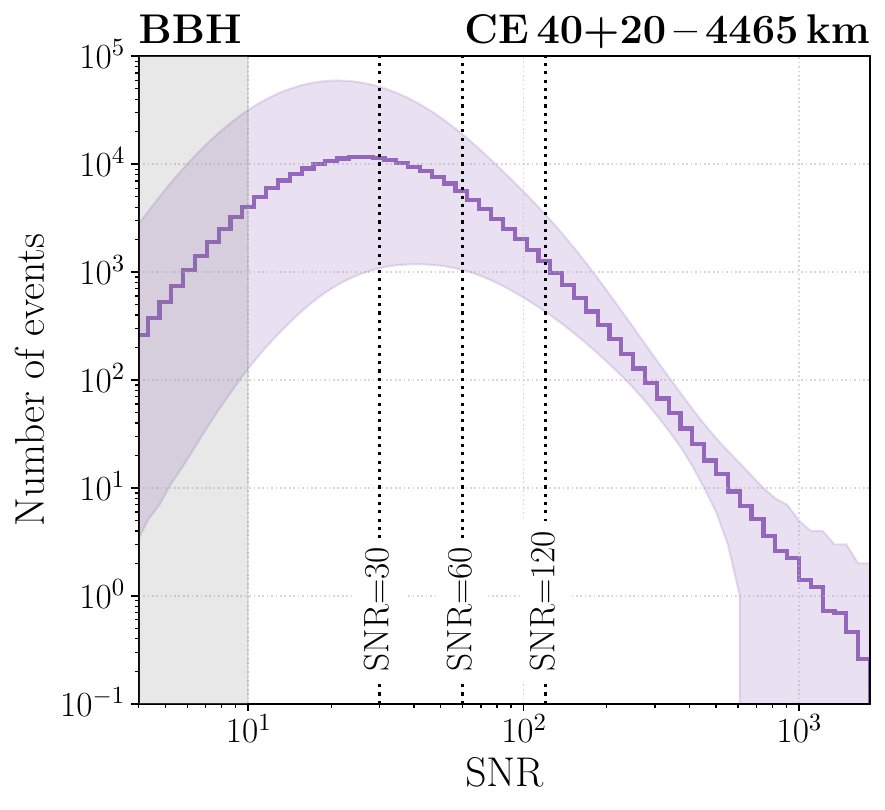}\\
        \caption{SNR distribution of 100 realizations of a 1\,yr BBH catalog consistent with the latest LVK results~\cite{LIGOScientific:2025pvj}, as observed by 2 CE detectors with a baseline of 4465\,km. The histogram represents the median SNR in each bin.
         Vertical dotted lines correspond to the three SNR values considered in this work, which are close to the median, upper 68\%, and upper 95\% percentiles of the SNR distribution. The gray shaded region marks the SNR detectability threshold of 10 used in our population analysis.}
        \label{fig:snr_distr_CE}
    \end{figure}

    \begin{figure}[tb]
        \centering
        \includegraphics[width=\linewidth]{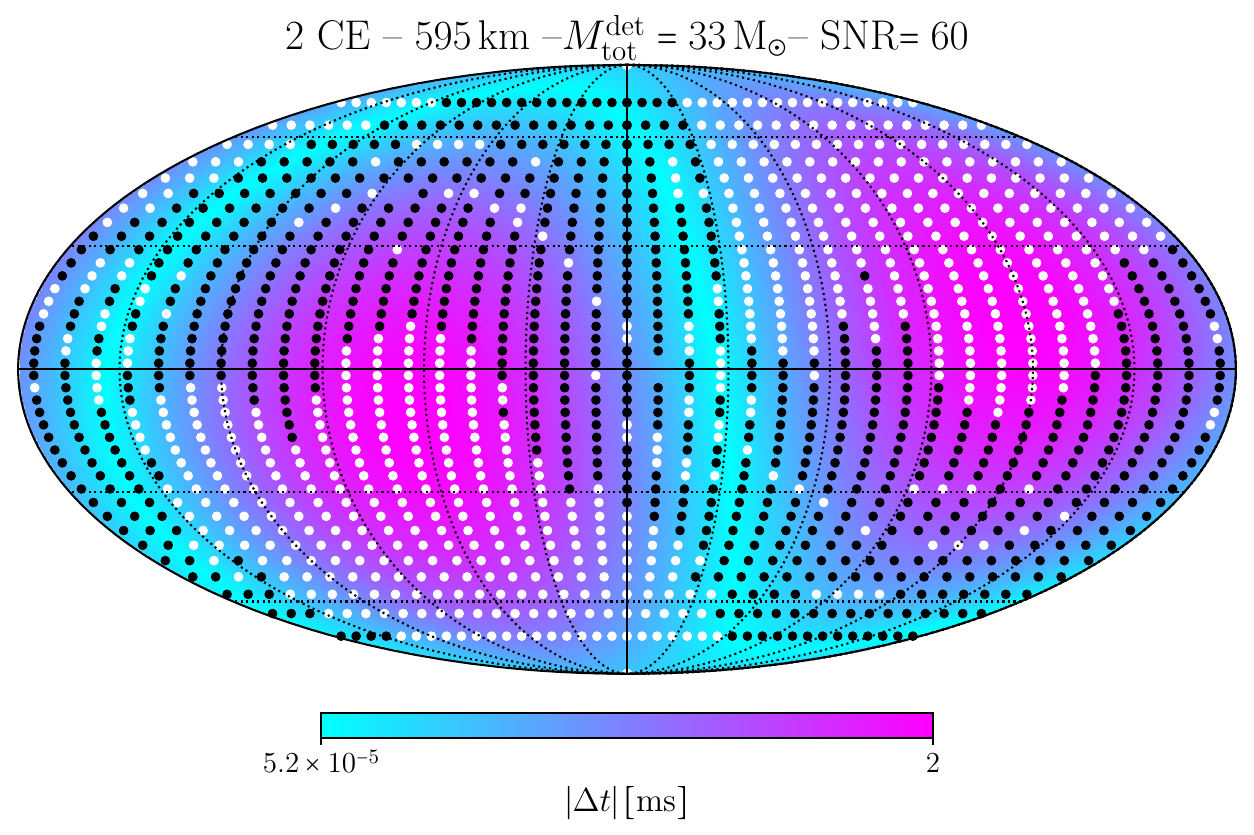}\\
        \includegraphics[width=\linewidth]{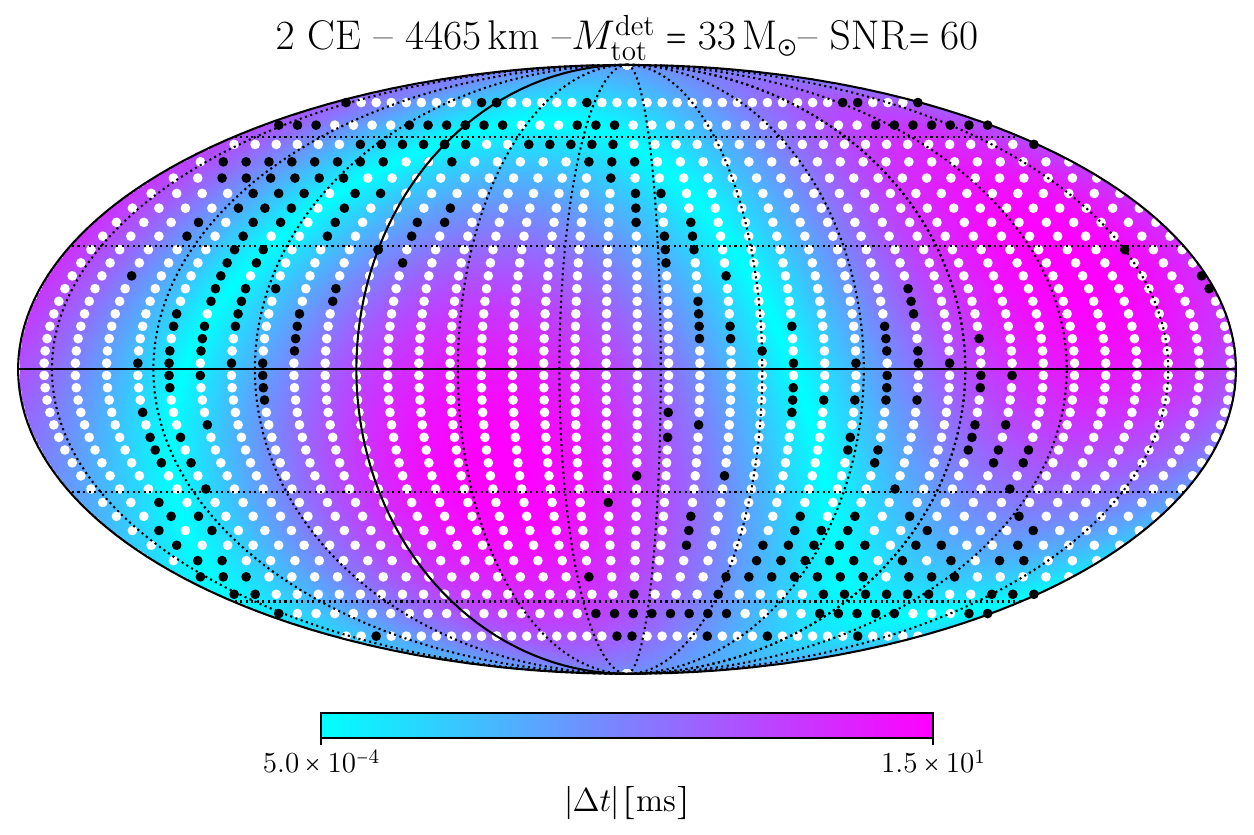}
        \caption{Mollweide projections of the absolute value of the time delay between the two CE detectors for the most distant (4465\,km, lower panel) and closest (595\,km, upper panel) configuration considered. We overplot points corresponding to the position of the events considered in our simulations. The black color denotes points for which the sky localization posterior has more than two modes for equal-mass, face-on injections with a detector-frame total mass $M_{\rm tot}^{\rm det}=33\,{\rm M}_\odot$ and SNR=60. 
        The white color denotes positions in which the sky localization for those injections has at most two modes.}
        \label{fig:mollview_time_delay_points}
    \end{figure}
    
    \item\label{item:reconstruction} For each detector configuration and SNR value we thus have a total of $6.4\times10^{4}$ binaries. %
    We analyze them with the \texttt{BAYESTAR} package using the \textsc{IMRPhenomXAS} waveform approximant~\cite{Pratten:2020fqn}, which models only the dominant $(\ell=2,\,|m|=2)$ emission mode of the signal, and adopt a uniform prior in comoving volume for the luminosity distance. We resort to the \texttt{ligo-skymap-stats} executable to compute, from each skymap produced by \texttt{BAYESTAR}, the 90\% localization area and volume, as well as the number of modes in the sky. This last step is achieved via a \emph{flood fill} algorithm (see e.g. Ref.~\cite{Foley_book_graphycs}). We then compute two other quantities from the skymap: the number of distinct peaks in the marginalized luminosity distance posterior and, for systems with a multimodal sky localization posterior, the area $A_{\partialvardoiint}$ of the smallest ellipse containing all the modes in the 90\% contour. The number of peaks is obtained through the \texttt{find\_peaks} function implemented in \href{https://github.com/scipy/scipy}{\texttt{scipy}}~\cite{2020SciPy-NMeth} by finding the peaks in the posterior with a prominence higher than 2\% of the maximum. For each peak, we then fit a Gaussian distribution to the corresponding mode, and count only the modes contributing to at least 15\% of the posterior area. It may be possible to increase the Gaussianity of the posterior by working in a different set of variables, see e.g.~\cite{Roulet:2022kot}. A more detailed investigation of this possibility is left for future work.
    The quantity $A_{\partialvardoiint}$ is computed by numerically finding the 2D Löwner--John ellipse containing all the pixels in the 90\% contour of the skymap through Khachiyan's ellipsoid minimization method~\cite{Grotschel1993}. For multimodal posteriors, this minimum ellipse $A_{\partialvardoiint}$ provides information which is complementary to the sky localization area. The latter measures the size of each mode irrespective of its position in the sky, while the former encodes information on the spread of the modes, penalizing posteriors with modes distant from each other and thus more difficult to follow up with a single electromagnetic telescope. 
    \item We repeat step~\ref{item:reconstruction} for a realistic population of BBH sources. In this case, we simulate a population corresponding to 1\,yr of mergers using the maximum likelihood hyperparameter sample from the latest LVK results~\cite{LIGOScientific:2025pvj}, and the same distributions used for generating the catalogs in step~\ref{item:snr_distr}. We restrict our study to systems with aligned spins, using again the \textsc{IMRPhenomXAS} waveform approximant. We analyze only systems that spend less than 5\,min in the detectors' sensitivity band, and impose a detectability threshold of 10 on the network SNR, and 4 on the SNR for each individual detector. 
    
\end{enumerate}

In Fig.~\ref{fig:mollview_time_delay_points}, we provide an example of the output of our analysis for equal-mass, face-on binaries with detector-frame total mass $M_{\rm tot}^{\rm det}=33\,{\rm M}_\odot$ observed by two CE detectors separated by 4465\,km and 595\,km and a network SNR of 60. In each panel, the color shows the absolute value of the time delay between the two detectors, and each point in the sky represents one of our simulations. Black points denote positions in which the sky localization posterior has more than two modes, white points denote positions in which the sky localization posterior is at most bimodal. One can immediately appreciate the impact of the longer baseline: for detectors separated by 4465\,km, only about 18\% of the injections show a posterior with more than 2 modes, while this number rises to about 51\% with detectors separated by 595\,km. Consistently, we also note how the black points broadly follow the regions with a short time delay. This is to be expected: when fixing the SNR, we are fixing the timing accuracy, hence the shorter the time delay, the less informative is the timing posterior, which provides the main contribution to the sky localization (see e.g.~\cite{Fairhurst:2009tc,Fairhurst:2010is}). Indeed, as shown in Ref.~\cite{Fairhurst:2009tc} in the Gaussian approximation to the posterior and considering only the timing information, the sky localization area scales as the inverse of the distance between the detectors.
This leads to the structure we observe in Fig.~\ref{fig:mollview_time_delay_points}, where most of the points with multimodal posterior in the lower panel fall in regions with time delay lower than the timing accuracy for the chosen event and SNR, $\sigma_t \simeq 9\times10^{-2}\,{\rm ms}$. The distribution of points with multimodal posteriors in the upper panel does not follow this simple trend, but it still traces the time-delay distribution. This explanation holds despite the fact that \texttt{BAYESTAR} accounts for information from the amplitude and phase of the signal along with the timing, highlighting the fact that timing really is the main driver of the sky localization accuracy~\cite{Fairhurst:2009tc}. 

\begin{figure*}[tbp]
    \centering
    \includegraphics[width=0.988\linewidth]{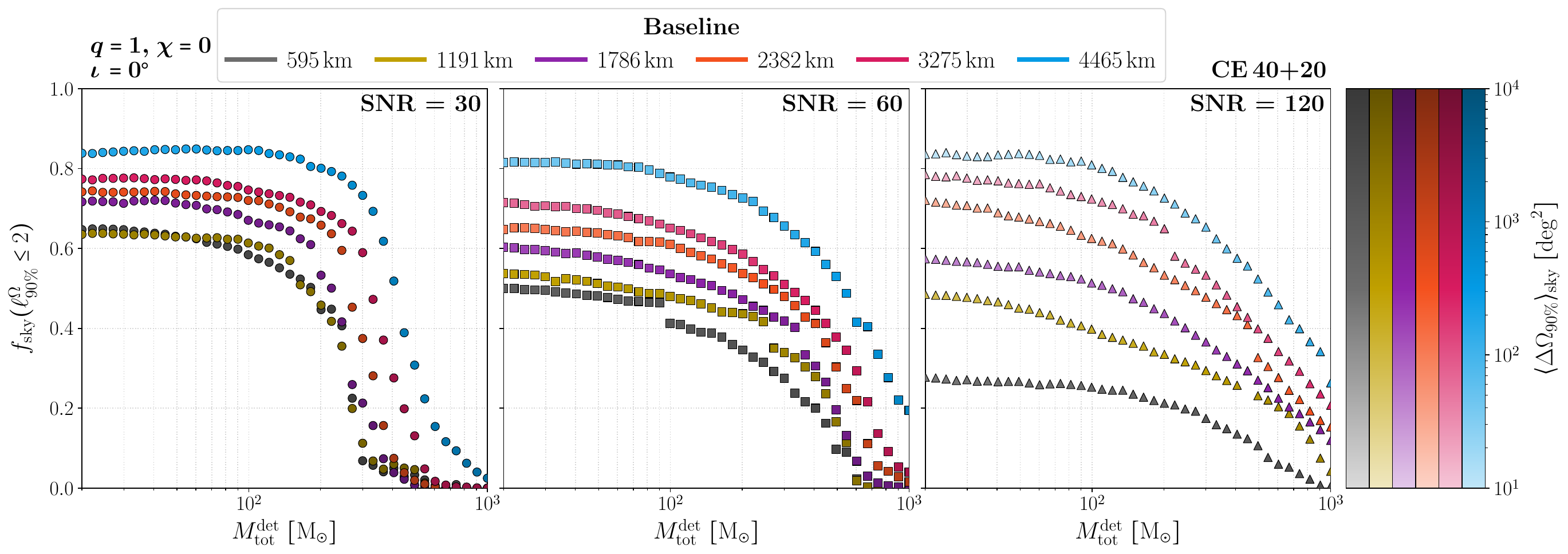}\\
    \includegraphics[width=0.988\linewidth]{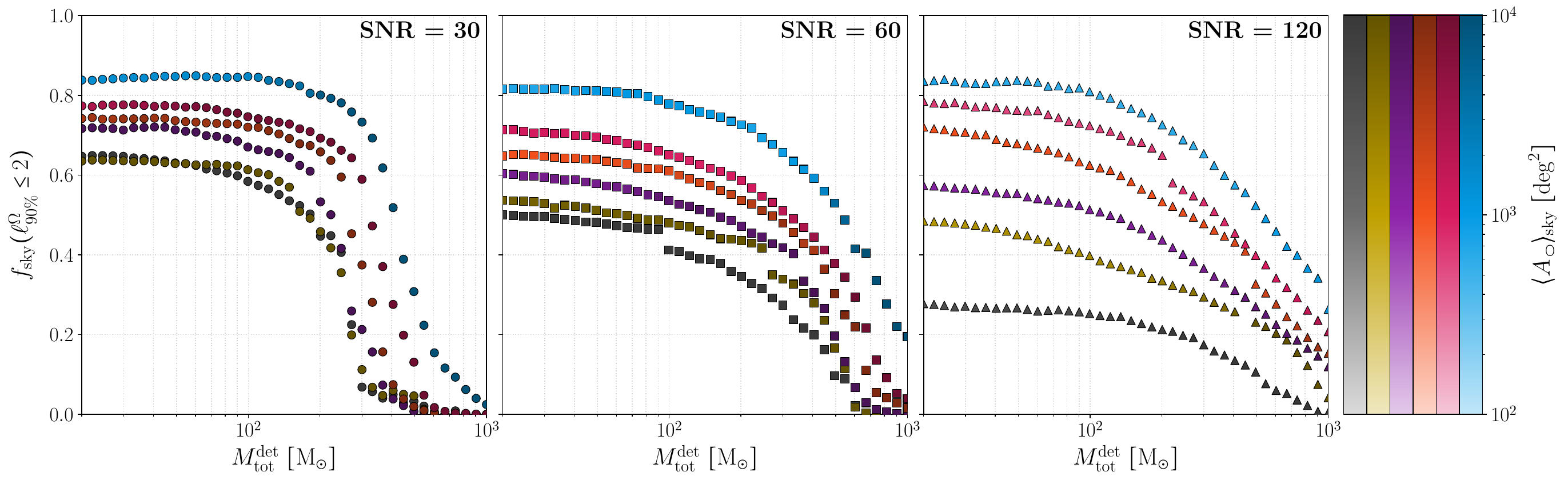} \\
    \includegraphics[width=\linewidth]{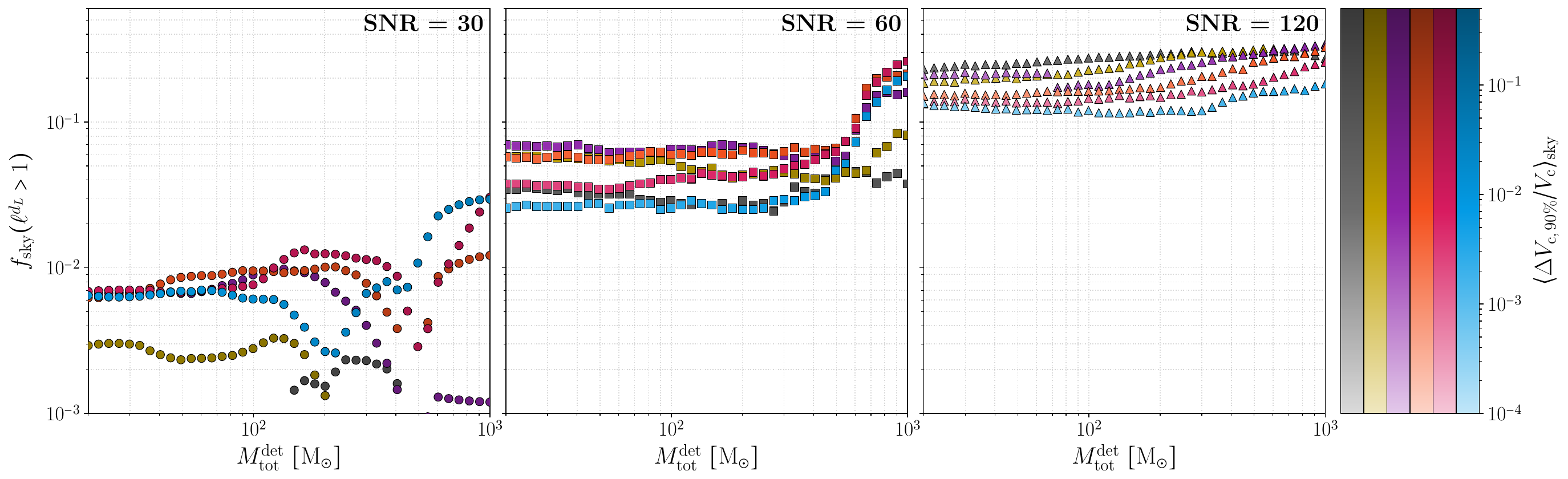}
    \caption{Upper panels: Fraction of the points in the sky exhibiting a sky localization posterior with at most two distinct modes as a function of the detector-frame total mass. Different colors correspond to different distances between CE detectors, while the shade of the points encodes the average 90\% sky localization area. Central panels: same as the upper panels, but with the shade showing the area of the smallest ellipse that encloses all the modes of multimodal posteriors. Lower panels: fraction of the points in the sky exhibiting a multimodal luminosity distance posterior as a function of the detector-frame total mass. The shades correspond to the average ratio between the 90\% inferred comoving volume and the total comoving volume at the redshift of the event. In each row, the left panel shows the results for events with ${\rm SNR}=30$, the central panel for ${\rm SNR}=60$, and the right panel for ${\rm SNR}=120$. All the results refer to equal-mass, face-on, nonspinning binaries.}
    \label{fig:fraction_multimodalities_baseline}
\end{figure*}

\section{Results}\label{sec:results}

In this section, we discuss our main sky localization results, including the impact of the baseline, the effect of the mass ratio and inclination of the sources, the improvements coming from the addition of other detectors to a two-CE network, and forecasts based on a population of sources consistent with current LVK observations.

\subsection{Impact of the baseline on the localization}\label{subsec:baseline_var}

\begin{table*}[tb]
    \centering
    {\setlength{\tabcolsep}{3.65pt}
    \begin{tabularx}{\linewidth}{c|c|ccc|ccc|ccc|ccc}
    \toprule\midrule
    \multirow{3}{*}{\begin{tabular}{c}
        \textbf{2 CE}\\
        \textbf{baseline}
    \end{tabular}} & \multirow{3}{*}{SNR} &
    \multicolumn{9}{c|}{percentage of systems} & \multicolumn{3}{c}{\multirow{2}{*}{$\langle\theta_{{\rm sep},\,90\%}\rangle$}} \\
    \cmidrule{3-11}
    & & \multicolumn{3}{c|}{in ZTF FoV} & \multicolumn{3}{c|}{in LSST FoV} & \multicolumn{3}{c|}{obs. by one hemisphere} & \\
    & & $20\,{\rm M}_\odot$ & $100\,{\rm M}_\odot$ & $400\,{\rm M}_\odot$ & $20\,{\rm M}_\odot$ & $100\,{\rm M}_\odot$ & $400\,{\rm M}_\odot$ & $20\,{\rm M}_\odot$ & $100\,{\rm M}_\odot$ & $400\,{\rm M}_\odot$ & $20\,{\rm M}_\odot$ & $100\,{\rm M}_\odot$ & $400\,{\rm M}_\odot$ \\
    \midrule
    \multirow{3}{*}{595\,km} & 30 & 2.1\% & 1.3\% & 0\% & 0.1\% & 0.1\% & 0\% & 15.3\% & 7.0\% & 0\% & 132$^\circ$ & 143$^\circ$ & 171$^\circ$ \\
    & 60 & 8.9\% & 5.1\% & 1.0\% & 0.8\% & 0.5\% & 0.3\% & 33.5\% & 23.8\% & 1.7\% & 113$^\circ$ & 123$^\circ$ & 156$^\circ$ \\
    & 120 & 19.8\% & 11.3\% & 1.6\% & 6.1\% & 1.6\% & 0.3\% & 51.2\% & 44.3\% & 15.5\% & 97$^\circ$ & 105$^\circ$ & 133$^\circ$ \\
    \midrule
    \multirow{3}{*}{1191\,km} & 30 & 5.1\% & 2.3\% & 0.6\% & 0.6\% & 0.1\% & 0.1\% & 31.9\% & 24.6\% & 1.3\% & 118$^\circ$ & 124$^\circ$ & 160$^\circ$ \\
    & 60 & 22.4\% & 13.9\% & 3.3\% & 5.0\% & 2.8\% & 0.3\% & 60.2\% & 55.2\% & 24.9\% & 91$^\circ$ & 98$^\circ$ & 127$^\circ$ \\
    & 120 & 33.4\% & 27.3\% & 8.6\% & 11.1\% & 8.1\% & 1.1\% & 71.1\% & 68.2\% & 48.7\% & 75$^\circ$ & 82$^\circ$ & 104$^\circ$ \\
    \midrule
    \multirow{3}{*}{1786\,km} & 30 & 9.5\% & 5.6\% & 1.5\% & 2.1\% & 0.9\% & 0.1\% & 54.0\% & 50.5\% & 5.1\% & 103$^\circ$ & 108$^\circ$ & 146$^\circ$ \\
    & 60 & 27.3\% & 19.9\% & 7.1\% & 6.5\% & 5.0\% & 1.3\% & 72.1\% & 68.7\% & 47.2\% & 79$^\circ$ & 86$^\circ$ & 109$^\circ$ \\
    & 120 & 44.8\% & 33.3\% & 19.1\% & 14.6\% & 10.0\% & 3.8\% & 80.0\% & 78.4\% & 68.0\% & 63$^\circ$ & 68$^\circ$ & 88$^\circ$ \\
    \midrule
    \multirow{3}{*}{2382\,km} & 30 & 11.2\% & 8.2\% & 2.8\% & 3.2\% & 1.7\% & 0.3\% & 69.6\% & 66.8\% & 14.8\% & 88$^\circ$ & 93$^\circ$ & 133$^\circ$ \\
    & 60 & 31.7\% & 24.5\% & 11.7\% & 8.2\% & 5.6\% & 2.6\% & 83.8\% & 81.6\% & 68.7\% & 66$^\circ$ & 71$^\circ$ & 94$^\circ$ \\
    & 120 & 63.3\% & 44.0\% & 26.1\% & 18.1\% & 11.0\% & 4.8\% & 90.2\% & 88.8\% & 82.4\% & 48$^\circ$ & 53$^\circ$ & 70$^\circ$ \\
    \midrule
    \multirow{3}{*}{3275\,km} & 30 & 17.2\% & 10.9\% & 4.7\% & 3.6\% & 3.1\% & 0.4\% & 75.2\% & 78.9\% & 44.7\% & 80$^\circ$ & 82$^\circ$ & 118$^\circ$ \\
    & 60 & 42.5\% & 31.2\% & 17.5\% & 11.5\% & 8.3\% & 3.8\% & 84.1\% & 83.0\% & 79.7\% & 59$^\circ$ & 64$^\circ$ & 86$^\circ$ \\
    & 120 & 71.0\% & 68.9\% & 34.1\% & 27.1\% & 21.2\% & 9.7\% & 93.4\% & 92.7\% & 88.6\% & 46$^\circ$ & 50$^\circ$ & 65$^\circ$ \\
    \midrule
    \multirow{3}{*}{4465\,km} & 30 & 23.8\% & 19.5\% & 9.1\% & 4.1\% & 3.0\% & 2.0\% & 86.7\% & 85.3\% & 64.6\% & 65$^\circ$ & 69$^\circ$ & 105$^\circ$ \\
    & 60 & 69.0\% & 43.2\% & 31.3\% & 19.2\% & 10.9\% & 7.3\% & 92.5\% & 92.1\% & 87.4\% & 45$^\circ$ & 49$^\circ$ & 70$^\circ$ \\
    & 120 & 77.2\% & 79.0\% & 52.3\% & 48.5\% & 27.6\% & 18.9\% & 95.4\% & 94.7\% & 92.7\% & 31$^\circ$ & 35$^\circ$ & 50$^\circ$ \\
    \midrule\bottomrule
    \end{tabularx}
    }
    \caption{For the different baselines of two CE instruments considered we report, from left to right: the fraction of detections whose localization area can be covered by a single pointing of ZTF and LSST, the fraction of the detections with sky localization such that it could be entirely observable by a single instrument in one hemisphere, and the average angular separation of the most distant points in the sky map. For each quantity and baseline, different columns refer to binaries with detector-frame total masses of $M_{\rm tot}^{\rm det} \simeq 20,\,100,\,400\,{\rm M}_\odot$ respectively, while different rows refer to the three chosen SNR values.}
    \label{tab:2ce_grid_res}
\end{table*}

We start by focusing on equal-mass systems and varying the baseline between the CE detectors. The results are shown in Fig.~\ref{fig:fraction_multimodalities_baseline}. In the upper row, for each mass value we report the fraction of the injections over the sky exhibiting a sky localization posterior with less than two modes at 90\% confidence level (c.l.), $f_{\rm sky}(\ell^\Omega_{90\%}\leq 2)$. Different colors correspond to different detector baselines, while the shading represents the average localization area across injections for each point. Columns correspond to different SNR values. 
As expected, we find the longest baseline to provide the best results, as less than 20\% of the systems present a sky localization posterior with more than two modes over a range of masses extending from $M_{\rm tot}^{\rm det}=20\,{\rm M}_\odot$ to $100\,{\rm M}_\odot$, irrespectively of the SNR. For the lowest SNR we consider, posteriors are at most bimodal for more than $\sim\!80\%$ of the sky up to $M_{\rm tot}^{\rm det}\sim200\,{\rm M}_\odot$, while this threshold in mass decreases for larger SNRs. In fact, as shown by the shading of the different points, when the SNR increases, the total localization area shrinks, and different modes separate into different patches of the sky. For lower SNRs, instead, the sky localization is less precise, and distinct peaks in the sky localization posterior merge together.
A similar trend is observed for the baselines of 3275\,km and 2382\,km when going from ${\rm SNR}=30$ to 120, albeit in this case only $\sim \!70\%$ and $\sim\!60\%$ of the injections up to $M_{\rm tot}^{\rm det}\sim 100\,{\rm M}_\odot$ show posteriors that are at most bimodal, respectively. For ${\rm SNR}=60$, we instead observe a larger degradation in $f_{\rm sky}(\ell^\Omega_{90\%}\leq 2)$. This can be traced to the competing effect of the overall localization improving thanks to the higher SNR, and different sky patches starting to separate. For shorter baselines, the results are qualitatively different. 
Decreasing the detector distance to 1786\,km gives comparable trends to the previous cases for ${\rm SNR}=30$ and 60, albeit lowering again the fraction of events with at most bimodal localization to $\sim\!50\%$. However, we observe a decrease in $f_{\rm sky}(\ell^\Omega_{90\%}\leq 2)$ for ${\rm SNR}=120$. This points to the fact that the posterior starts to be more degenerate due to the short baseline, not allowing for the selection of a single mode at 90\% c.l. even with a large SNR. 
For detector distances of 1191\,km and 595\,km we find a consistent result for the ${\rm SNR}=30$ case, with $\sim\!60\%$ of the simulations resulting in localization posteriors at most bimodal up to $M_{\rm tot}^{\rm det}\sim 100\,{\rm M}_\odot$. The drop in this fraction is more evident than in the previous cases for higher SNRs, in particular for the shortest baseline we consider. In this scenario, for ${\rm SNR}=120$, only about 25\% of the systems have at most two modes, which points to a less informative and more degenerate posterior compared to the longer baselines.
Regarding the average localization area, as expected, we find lighter systems to provide better results for all the baselines, given the longer time they spend in the detectors' sensitivity band. For the lowest SNR value, we do not find average localization areas smaller than $\sim\!800\,{\rm deg}^2$.
Increasing the SNR improves the results for the longer baselines, with average localization areas of a few tens (hundreds) of square degrees for ${\rm SNR}=120$ (60). For the shortest baseline, localization areas below $\Delta\Omega_{90\%} \sim100\,{\rm deg}^2$ are hardly achievable, even at high SNR.

In the second row of Fig.~\ref{fig:fraction_multimodalities_baseline} we show once again the fraction $f_{\rm sky}(\ell^\Omega_{90\%}\leq 2)$ as a function of mass, but this time the shading of the points corresponds to the average area of the smallest ellipse containing all the patches in the sky localization posterior. In comparison with the results in the first row, we find that the improvement in the area of this ellipse with increasing SNR is less prominent than the improvement in the total sky localization area, even for the longest baselines. 
This is because a larger SNR causes the localization area of each individual mode in the sky to shrink, but the modes will remain sufficiently far apart from each other. To further elaborate on this point, in the rightmost column of Table~\ref{tab:2ce_grid_res} we report the average angular separation between modes for multimodal posteriors $\langle\theta_{\rm sep, 90\%}\rangle$ as a function of baseline and SNR for three representative values of the mass. For a given baseline and SNR, different modes for multimodal posteriors are, on average, further apart from each other with increasing mass, as expected. In particular, for massive binaries, we find that we cannot achieve an average separation between modes less than $50^\circ$, even with $\rm{SNR}=120$ and the longest baseline we consider.
In this situation, an early electromagnetic follow-up looking for fast transients on the timescale of a few seconds would require more than one instrument, even if the total sky localization area is nominally small. For comparison, a GRB-dedicated telescope such as the Neil Gehrels {\it Swift} Observatory (\textit{Swift})~\cite{SwiftScience:2004ykd} Burst Alert Telescope (BAT)~\cite{Barthelmy:2005hs} has a slewing time between two positions separated by $\theta_{{\rm sep}}$ degrees in the sky of $t_{\rm slew} = (25 +2\,\theta_{{\rm sep}})\,{\rm s}$~\cite{Eyles-Ferris:2024bkl}. 

Focusing instead on EM follow-up campaigns targeting transients evolving on timescales of days to weeks since merger, we can consider instruments such as the Zwicky Transient Facility (ZTF), with a field of view (FoV) of $\sim\!47\,{\rm deg}^2$~\cite{Graham:2019qsw} and some observing time devoted to follow-ups of GW events~\cite{Ahumada:2024qpr}, and the Vera Rubin Large Synoptic Survey Telescope (LSST)~\cite{LSST:2008ijt}, with a FoV of $\sim\!9.6\,{\rm deg}^2$ and a target of opportunity (ToO) program which includes GWs~\cite{Andreoni:2024pkp}. The last is comparable to the FoV of DSA-2000, a future radio telescope spanning the 0.7-2\,GHz range, foreseen to have an instantaneous FoV of $10.6$\,deg$^2$~\cite{2024AAS...24326102P}. In these cases, maximizing the fraction of GW localization area observed with a single telescope pointing is key for minimizing the time spent on a given GW source. In Table~\ref{tab:2ce_grid_res} we thus report the fraction of points in the sky where the reconstruction of the sky localization (including all the patches for multimodal posteriors) falls entirely in the FoV of ZTF or LSST (approximated as a square projected onto the sky), respectively. Results are once again shown as a function of baseline and SNR for three representative masses. Averaging over all the masses and positions in the sky, we find that, for the longest baseline we consider, a fraction of $\sim\!15\%$ (42\%, 67\%) of the injections would be observable with a single ZTF pointing at ${\rm SNR}=30$ (60, 120), with lighter binaries being easier to follow up, as one can see from Table~\ref{tab:2ce_grid_res}. These values reduce to $\sim\!6\%$ (20\%, 40\%) for a baseline of 2382\,km and $\sim\!1\%$ (4\%, 9\%) for a baseline of 595\,km. For the remaining fraction of events, more than one pointing from a single telescope will be needed, resulting in a higher demand for observing time at that instrument to cover the localization area of a single GW event. Moreover, we can take into account that, depending on the kind of counterpart one is after, the considered telescopes would not be able to follow up candidates at high redshift. Limiting the analysis to events within $z=0.5$, we find that the fraction of injections observable with a single ZTF pointing at ${\rm SNR}=30$ (60, 120) for the longest baseline decreases to $\sim\!4\%$ (20\%, 41\%), for the baseline of 2382\,km to $\sim\!0.4\%$ (3\%, 18\%), and for the shortest baseline to $\sim\!0.1\%$ (1\%, 2\%).

In addition, one should also consider that a single telescope on the ground cannot cover the whole sky. Hence, in Table~\ref{tab:2ce_grid_res} we also show for some representative masses the percentage of points in the sky for which all of the modes of the sky localization posterior fall in a region of the sky observable with a single instrument in one hemisphere. Given the footprint of the ZTF instrument~\cite{2019PASP..131a8003M}, we assume this region to correspond to declinations from $90^\circ$ to $-30^\circ$ for the northern hemisphere and from $30^\circ$ to $-90^\circ$ for the southern hemisphere. Notice that we include both unimodal and multimodal sky localization posteriors. Averaging over all the mass values and sky positions considered, for the longest baseline we find that a fraction of $\sim\!28\%$ (12\%, 7\%) of the injections at ${\rm SNR}=30$ (60, 120) would be localized with sky patches in different hemispheres, in which case different instruments would be required to search for a potential electromagnetic emission. These numbers increase to $\sim\!52\%$ (30\%, 16\%) for a baseline of 2382\,km and $\sim\!94\%$ (83\%, 67\%) for the shortest baseline, with lighter binaries again providing the better constrained localization regions; see Table~\ref{tab:2ce_grid_res}. For the longest baseline, we find that events resulting in posteriors falling in different hemispheres for the highest SNR value are preferentially for events close to the blind spot of one of the detectors or close to the cut we impose on the localization in a single hemisphere of $\pm30^\circ$.

Finally, we focus on the inference of the luminosity distance $d_L$. As recently discussed in Ref.~\cite{Santoliquido:2025aiq}, for short signals the marginal posterior in $d_L$ might show a multimodal structure for a network of two instruments due to correlations with a multimodal sky localization. In the bottom row of Fig.~\ref{fig:fraction_multimodalities_baseline}, we show the fraction of injections over the sky that exhibit a multimodal luminosity distance posterior $f_{\rm sky}(\ell^{d_L}>1)$ as a function of mass, according to the procedure outlined in Sec.~\ref{subsec:loc_method_injs}.
The shading of the points represents the average inferred comoving volume at 90\% c.l. over the total comoving volume at the redshift of the event. We observe that increasing the SNR leads to more multimodal posteriors, with ${\cal O}(1\%)$ ($\lesssim10\%$, $\lesssim20\%$) of the simulated injections at different masses having a multimodal structure in $d_L$ for ${\rm SNR}=30$ (${\rm SNR}=60$, ${\rm SNR}=120$). Even though the trends are less clear compared to the sky localization for the lowest SNR value, the fraction of multimodal posteriors tends to increase with increasing mass. Both of these behaviors can be traced to the degeneracy between the localization and distance posteriors: as observed in the upper panels of Fig.~\ref{fig:fraction_multimodalities_baseline}, the fraction of events with multimodal posteriors increases with increasing SNR and mass. In any case, despite presenting a more multimodal structure in the posteriors, we still find the injections with the highest SNR to result in the lowest localization volumes. A note of caution is needed here: while in this part of the analysis we focus on face-on, nonspinning, equal-mass binaries, for which only the dominant emission mode is present, in general employing a waveform model including information from higher-order harmonics and spin precession would improve the reconstruction of the luminosity distance by breaking its degeneracy with the inclination (see e.g.~\cite{Ng:2022vbz,Mascioli:2025cnw}). In this sense, our results should be regarded as pessimistic. Indeed, in Appendix~\ref{app:highermodes} we study how the inclusion of subdominant harmonics in the analysis is able to alleviate the multimodality issue in the $d_L$ posterior, especially for high-SNR events.

\subsection{Impact of the mass ratio}\label{subsec:mass_ratio_var}

\begin{figure*}[tb]
    \centering
    \includegraphics[width=0.988\linewidth]{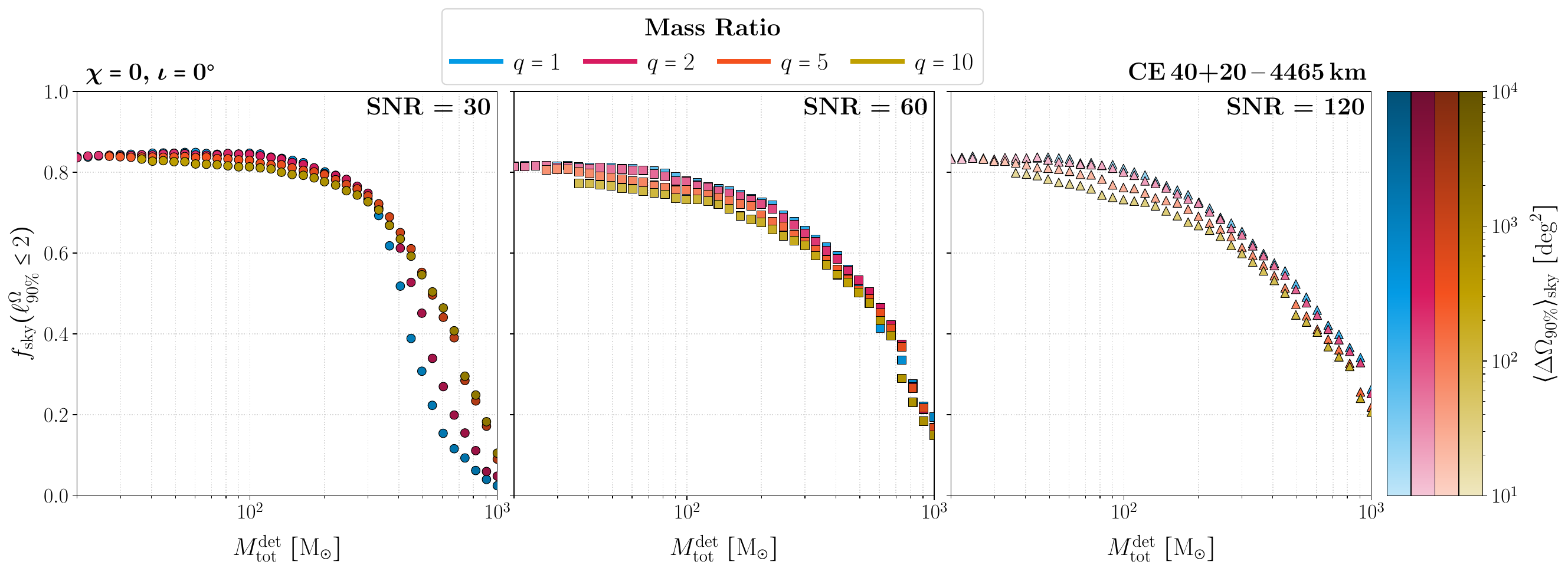}\\
    \includegraphics[width=\linewidth]{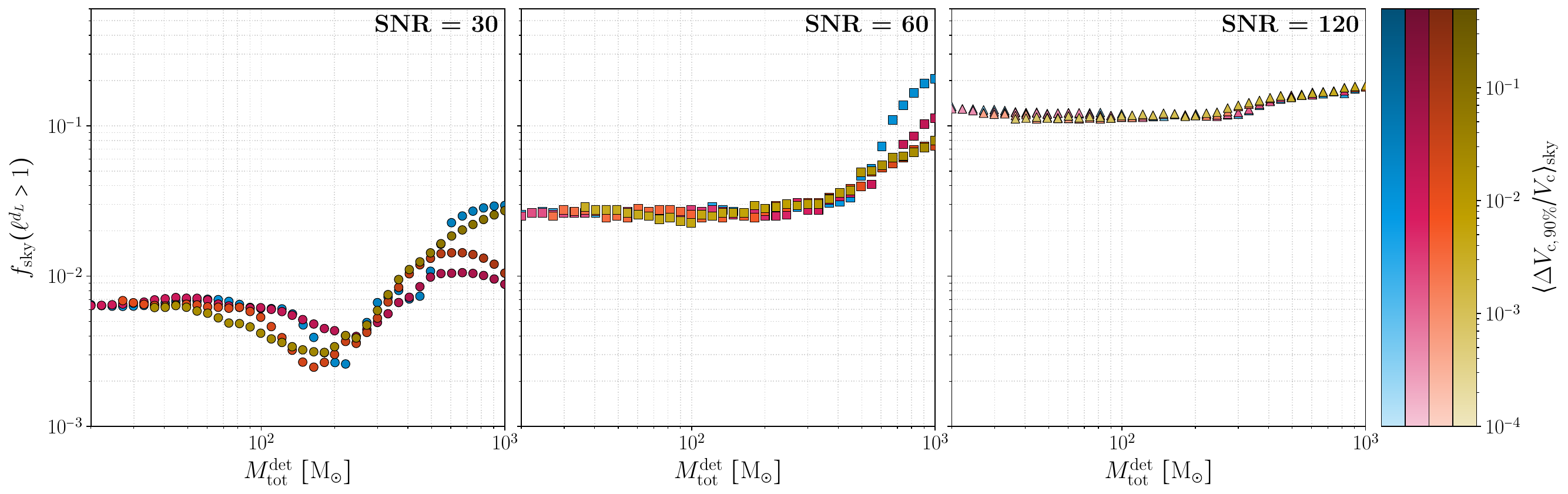}
    \caption{Same as in the top and bottom rows of Fig.~\ref{fig:fraction_multimodalities_baseline} for a fixed baseline of 4465\,km. Different colors correspond to different values of the binary mass ratio. For each mass ratio, we report only the values of the mass for which the signal spends less than 5\,min in the detectors' band. Results are shown for face-on, nonspinning binaries.
    }
    \label{fig:fraction_multimodalities_q}
\end{figure*}

\begin{figure*}[tb]
    \centering
    \includegraphics[width=0.988\linewidth]{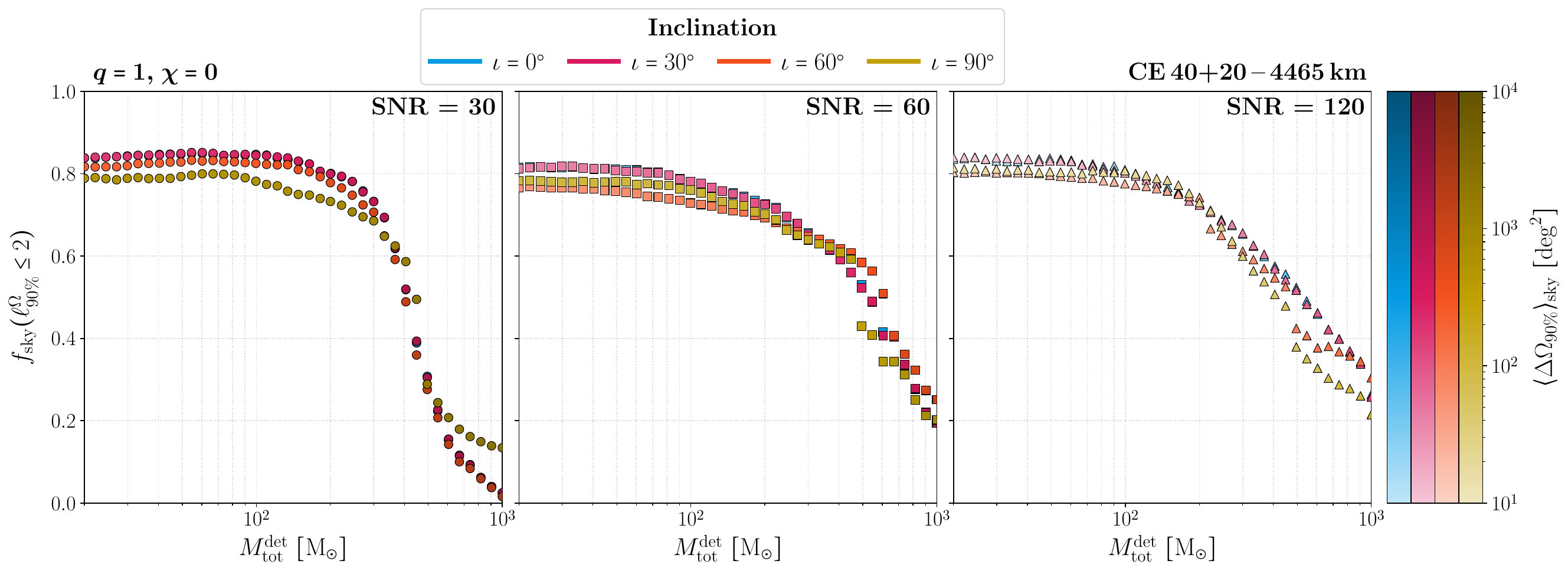}\\
    \includegraphics[width=\linewidth]{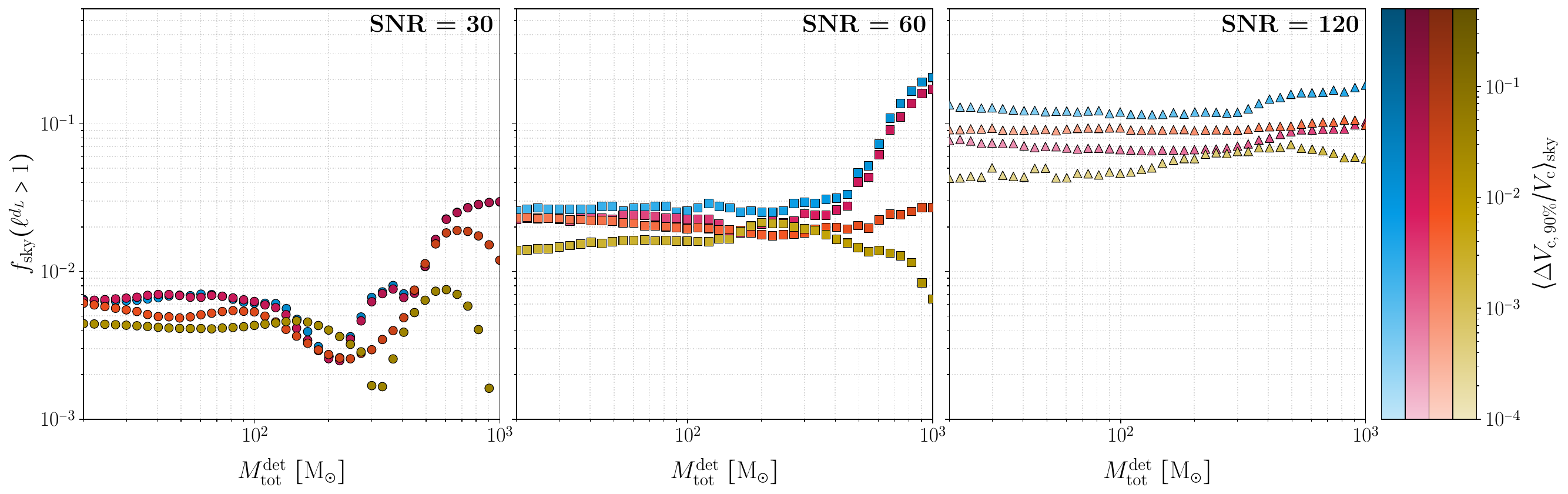}
    \caption{Same as in Fig.~\ref{fig:fraction_multimodalities_q} but fixing the mass ratio to $q=1$ and with different colors corresponding to different binary inclinations for nonspinning systems.}
    \label{fig:fraction_multimodalities_inclination}
\end{figure*}

In the previous section, we focused on equal-mass events observed with different baselines. Here, we study the impact of varying the mass ratio. For fixed total mass, systems with larger mass ratio spend more time in the detectors' frequency band [see Eq.~\eqref{eq:time_to_coal}]. In Fig.~\ref{fig:fraction_multimodalities_q}, we plot once again the fractions $f_{\rm sky}(\ell^\Omega_{90\%}\leq 2)$ and $f_{\rm sky}(\ell^{d_L}>1)$ as a function of the detector-frame total mass for the three SNRs we consider. However, this time we fix the CE baseline to the longest one in our analysis, while the colors correspond to four representative mass-ratio values $q=m_1/m_2 = 1,\, 2,\, 5,\, 10$. The shades represent the average inferred sky area (upper row) and the average ratio between inferred and total comoving volume at the redshift of the event (bottom row). We find that the sky localization and luminosity distance posteriors are not significantly affected by changing the mass ratio. For the lowest SNR, larger mass ratios improve the sky localization for more massive systems, which can be explained by the longer observation time. 
Given these results, we expect our findings for face-on binaries from Sec.~\ref{subsec:baseline_var} to hold also for asymmetric binaries. 

\subsection{Impact of the inclination}\label{subsec:inclination_var}

In this section, we explore the impact of the inclination angle $\iota$ between the binary's orbital angular momentum and the line of sight. This parameter controls how the two polarizations of the GW signal mix: for face-on binaries ($\iota = 0^\circ$), as those considered in the previous sections, the amplitudes of the two polarizations are equal for the fundamental emission mode, while for edge-on systems ($\iota = 90^\circ$) only one of the polarization modes survives~\cite{Maggiore:2007ulw}. Thus, disentangling the two polarizations might be more or less difficult depending on the value of $\iota$, in turn impacting the reconstruction of the distance, which is highly degenerate with the inclination.

In Fig.~\ref{fig:fraction_multimodalities_inclination} we plot the same quantities as in Fig.~\ref{fig:fraction_multimodalities_q}, once again fixing the CE baseline to be the longest considered in Sec.~\ref{subsec:baseline_var}. However, this time we focus on equal-mass binaries and consider four representative values of the inclination angle ($\iota=0^\circ,\,30^\circ,\,60^\circ,\,90^\circ$). 
From the upper panels, we notice how the impact of the inclination on the sky localization is marginal, with a slightly worse reconstruction for $\iota=60^\circ$ and $90^\circ$. This can be partially explained by the fact that, as the binary gets closer to an edge-on configuration, the amplitude of the $\times$ polarization of the signal decreases, and the dependence of the strain on the antenna pattern functions (and thus the sky position) becomes milder. For the luminosity distance, we observe the expected trend: as the inclination increases, the marginal posterior for $d_L$ becomes less and less degenerate with the one for $\iota$, leading to a weaker multimodal structure and to a mild improvement in the inference of the comoving volume for all the SNRs considered. Moreover, we find that binary configurations closer to the edge-on case result in an improved inference of the luminosity distance for binaries with high detector-frame masses. This can again be traced to the lesser degeneracy with the inclination posterior, which can be more severe for short signals. 

\subsection{Networks with additional detectors}\label{subsec:extended_net}

\begin{figure}[tb]
    \includegraphics[width=\columnwidth]{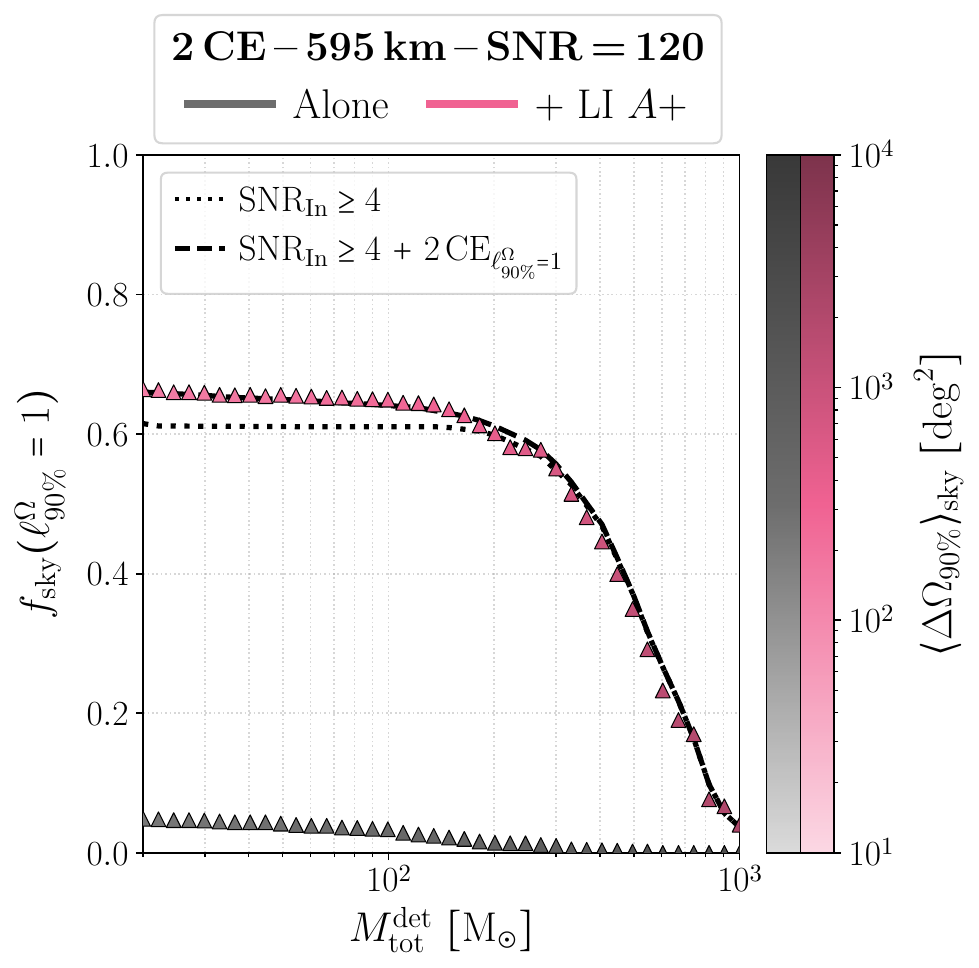} \caption{Fraction of the points in the sky exhibiting a unimodal sky localization posterior as a function of the detector frame mass (triangles). Different colors correspond to either a network of two CE detectors alone (gray triangles) or a three-detector network including also LIGO-India at $A+$ sensitivity (pink triangles). The shading encodes the average 90\% sky localization area. Results are shown for the shortest CE baseline and highest SNR considered in Fig.~\ref{fig:fraction_multimodalities_baseline}. For each mass, we also show the fraction of events with ${\rm SNR}\geq4$ in LIGO-India (dotted line), and the same fraction plus the events with unimodal localization with the 2 CE network not detected by LIGO-India (dashed line). All results refer to equal-mass, nonspinning binaries with face-on orientation.}
    \label{fig:fraction_multimodalities_india}
\end{figure}

In all the previous analyses, we only considered a detector network composed of two CE instruments. However, when at least a third detector is added to the network, multimodalities in the sky localization posterior are expected to disappear for events that are detected by all three instruments~\cite{Fairhurst:2009tc,Wen:2010cr,Fairhurst:2010is,KAGRA:2013rdx}. Strictly speaking, the event has to be detectable but not necessarily detected in all instruments for localization purposes, as was the case for GW170817~\cite{LIGOScientific:2017vwq}. However, for simplicity, we will not consider this scenario in our analysis. 

In Fig.~\ref{fig:fraction_multimodalities_india}, we thus select the configuration giving rise to the largest number of events with a multimodal structure in the posteriors from Fig.~\ref{fig:fraction_multimodalities_baseline}, namely the shortest baseline of 595\,km with ${\rm SNR}=120$, and repeat the analysis adding one LIGO-India detector with the target sensitivity for the $A+$ design. We rescale the distance to fix the SNR to 120 with the two CE detectors, 
and include the LIGO-India instrument in the analysis for events with ${\rm SNR}\geq4$ in this detector. We show the fraction of events with a unimodal sky localization posterior $f_{\rm sky}(\ell^\Omega_{90\%}=1)$ as a function of the total detector-frame mass, considering either the 2-CE network alone or with the inclusion of LIGO-India. We find that adding LIGO-India to the network, even at $A+$ sensitivity, increases the fraction of events with unimodal posterior significantly, from $\lesssim 10\%$ to more than 60\%. This number corresponds exactly to the events that are detected (even with modest SNR) by all three instruments in the network. To further prove this point, in Fig.~\ref{fig:fraction_multimodalities_india} we also show, for each mass: (\emph{i}) the fraction of events detected by LIGO-India with ${\rm SNR}\geq4$ (dotted line), and (\emph{ii}) the same quantity plus the fraction of events not detected by LIGO-India which are {\em already} showing a unimodal posterior with the two CE detectors alone (dashed line).
The latter curve matches the fraction of events with unimodal posteriors in the three-detector network (pink triangles) almost exactly. 
Moreover, the average localization area improves by about 60\% for masses $M_{\rm tot}\lesssim200\,{\rm M}_\odot$ when we include LIGO-India, thanks to the selection of a single mode in the sky. It is remarkable that a third detector, even with SNR as low as 4, can still contribute so significantly to the sky localization. This is not different from the case of GW170817, in which the operation of the Virgo interferometer along with the two LIGO instruments allowed for accurate localization of the source~\cite{LIGOScientific:2017vwq}.

\begin{figure*}[tb]
    \includegraphics[width=.98\linewidth]{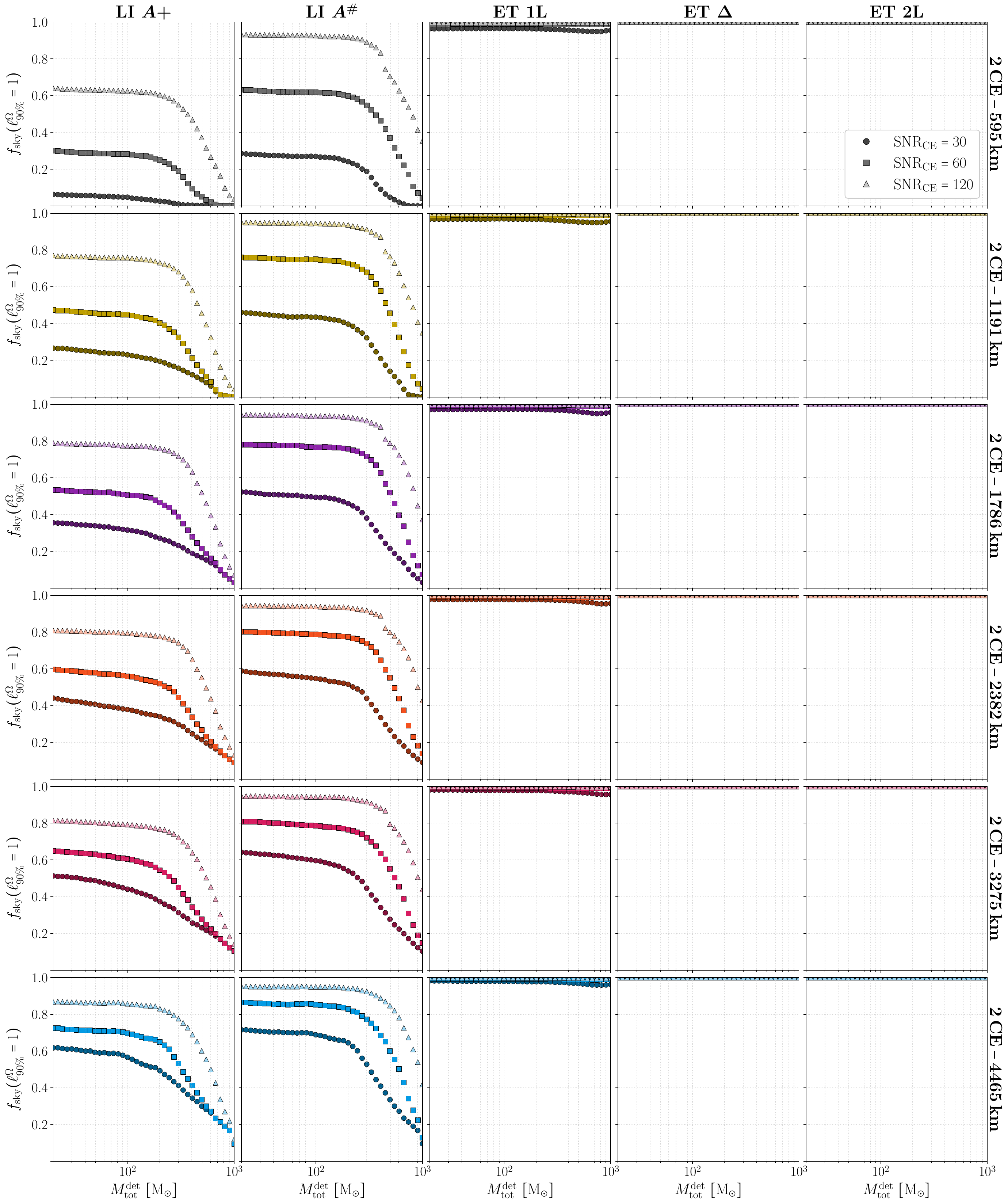} 
    \caption{Fraction of points in the sky exhibiting a unimodal sky localization posterior as a function of detector-frame mass for combinations of two CE interferometers with varying baseline (different rows) along with additional detectors in the network (different columns). We show results for the three SNRs considered with the two CE detectors:  ${\rm SNR}=30$ (dark-colored circles),  ${\rm SNR}=60$ (medium-colored squares), and  ${\rm SNR}=120$ (light-colored triangles). We consider equal-mass, nonspinning binaries with face-on orientation. %
    }
    \label{fig:fraction_unimodal_sky_deltaOmega_networks}
\end{figure*}

The results in Fig.~\ref{fig:fraction_multimodalities_india} justify a further approximation when adding a third detector to the network: we can safely assume that the multimodalities in the sky localization disappear for events observed by three detectors. Hence, from the same injections with a 2-CE network reported in Sec.~\ref{subsec:baseline_var}, we can compute the fraction of events with multimodal posteriors with different networks of more than two detectors by simply imposing {\em detectability} in the other detectors included in the network. In Fig.~\ref{fig:fraction_unimodal_sky_deltaOmega_networks} we show results for networks including 2 CE detectors with varying baselines (different rows) along with either LIGO-India with $A+$ or $A^\#$ sensitivity, a single L-shaped ET detector, ET in its triangular configuration, or 2 L-shaped misaligned ET detectors (different columns). The shape of the points (circles, squares, or triangles) represents different SNR values achieved with the two CE detectors. We find that adding a third detector to the network improves the results dramatically. Even a single LIGO-India detector with $A+$ sensitivity can boost the number of detections with unimodal posteriors significantly, effectively making the results for the highest SNR only mildly dependent on the CE baseline, with the only exception of the shortest baseline (top row). An important caveat is that when we consider lower SNR values in the 2 CE detectors, only a small fraction of events would be detected by LIGO-India with $A+$ sensitivity, making the CE baseline more important. Increasing LIGO-India's sensitivity to $A^\#$, most of the injections feature a unimodal structure for the highest SNR case and for masses up to $M_{\rm tot}\simeq300\,{\rm M}_\odot$. Even for the ${\rm SNR}=60$ case, more than 60\% of the injections have a unimodal localization for masses $M_{\rm tot}\lesssim300\,{\rm M}_\odot$, regardless of the CE baseline considered. The drop for higher mass values is a consequence of the smaller frequency band of the instruments. These results highlight the importance of having LIGO-India operating along with XG detectors for localization purposes.

When we consider combinations of two CE detectors with any of the ET configurations, we find that essentially none of the events would have a multimodal sky localization posterior, regardless of the CE baseline. The only mild exception is for ${\rm SNR}=30$ with the two CE detectors plus a single L-shaped ET, in which case a small fraction of $\sim\!2\%$ of the systems at high masses would feature a multimodal sky localization region. These findings strengthen the case for a global network of XG detectors.  

\subsection{Analysis of a realistic population of sources}\label{subsec:pop_analysis}

\begin{figure*}[tb]
    \centering
    \begin{tabular}{ccc}
        \includegraphics[width=.32\linewidth]{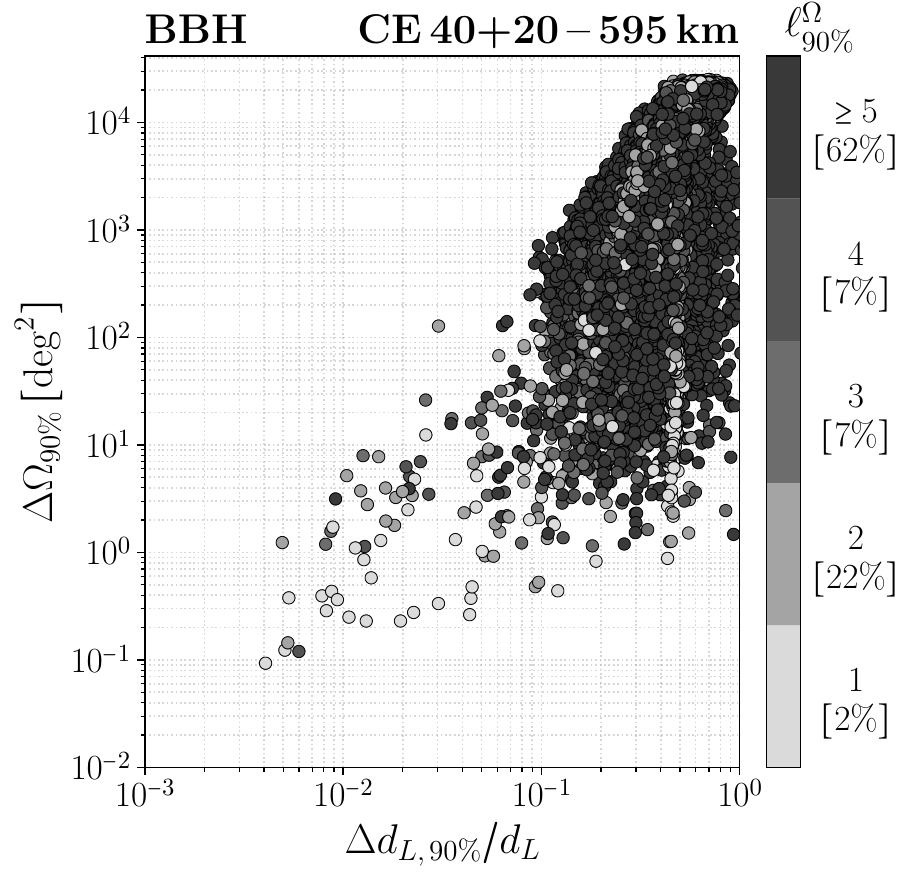} & \includegraphics[width=.32\linewidth]{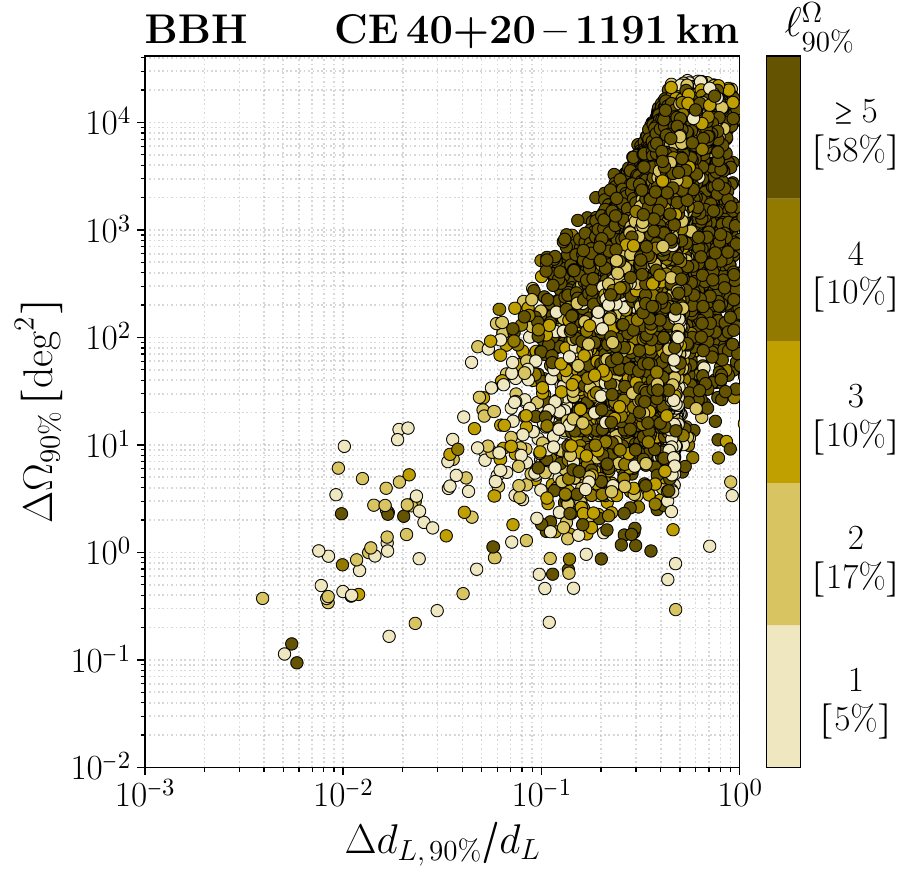} & \includegraphics[width=.32\linewidth]{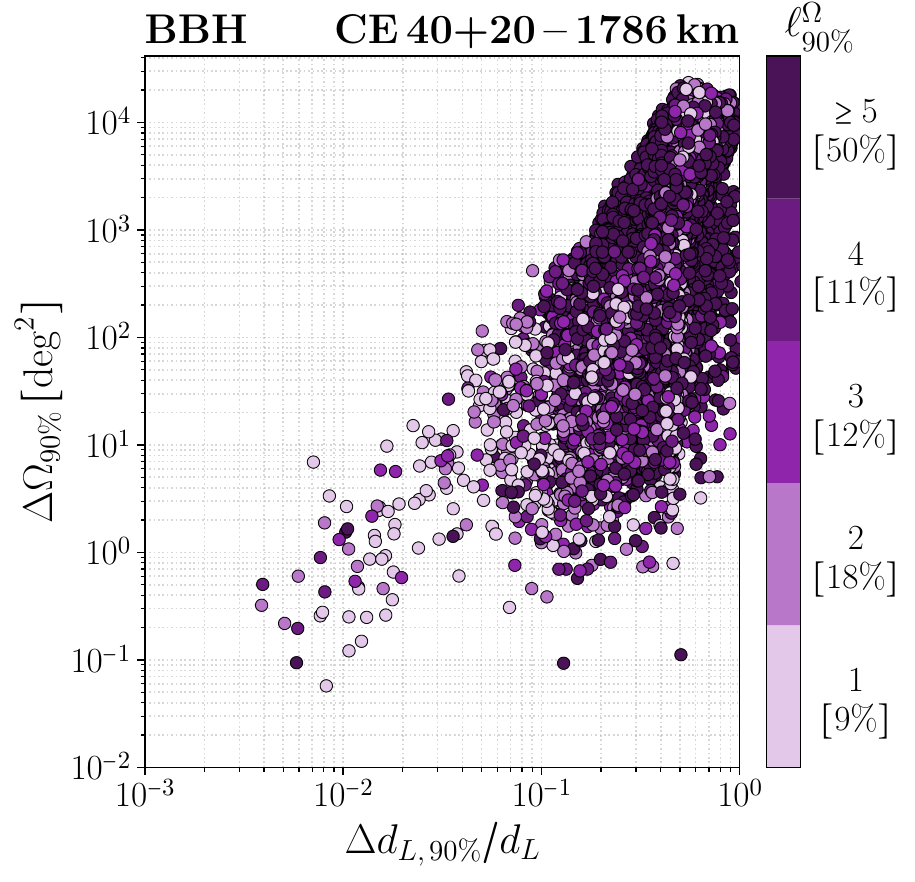} \\
        \includegraphics[width=.32\linewidth]{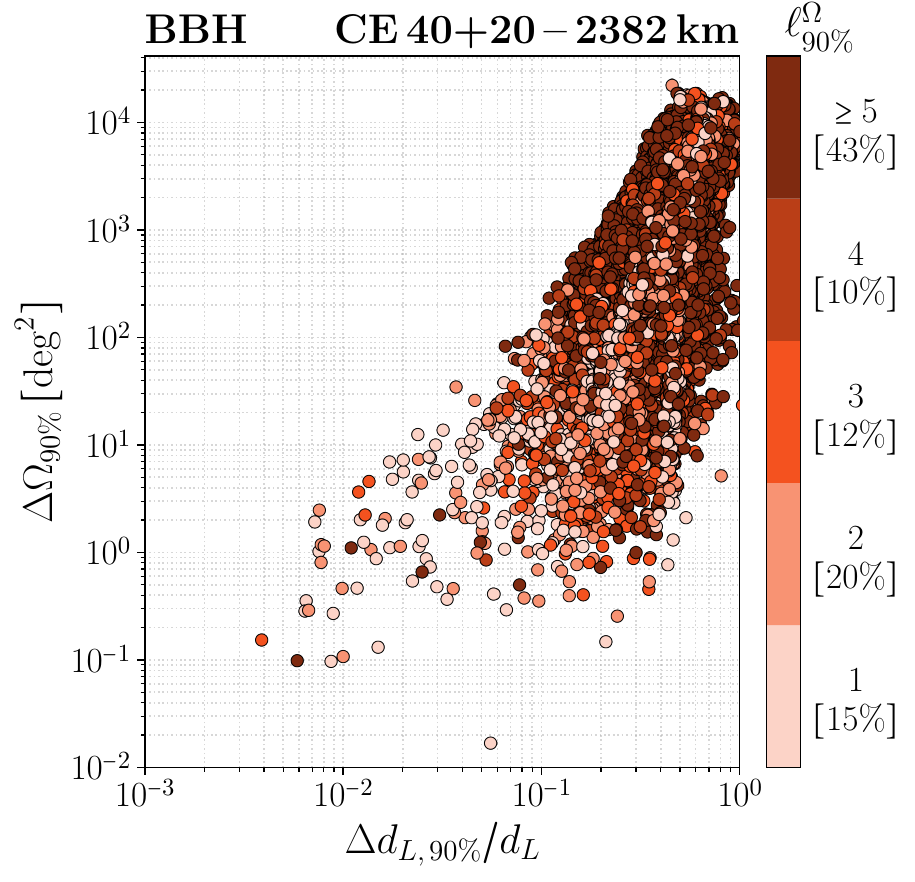} & \includegraphics[width=.32\linewidth]{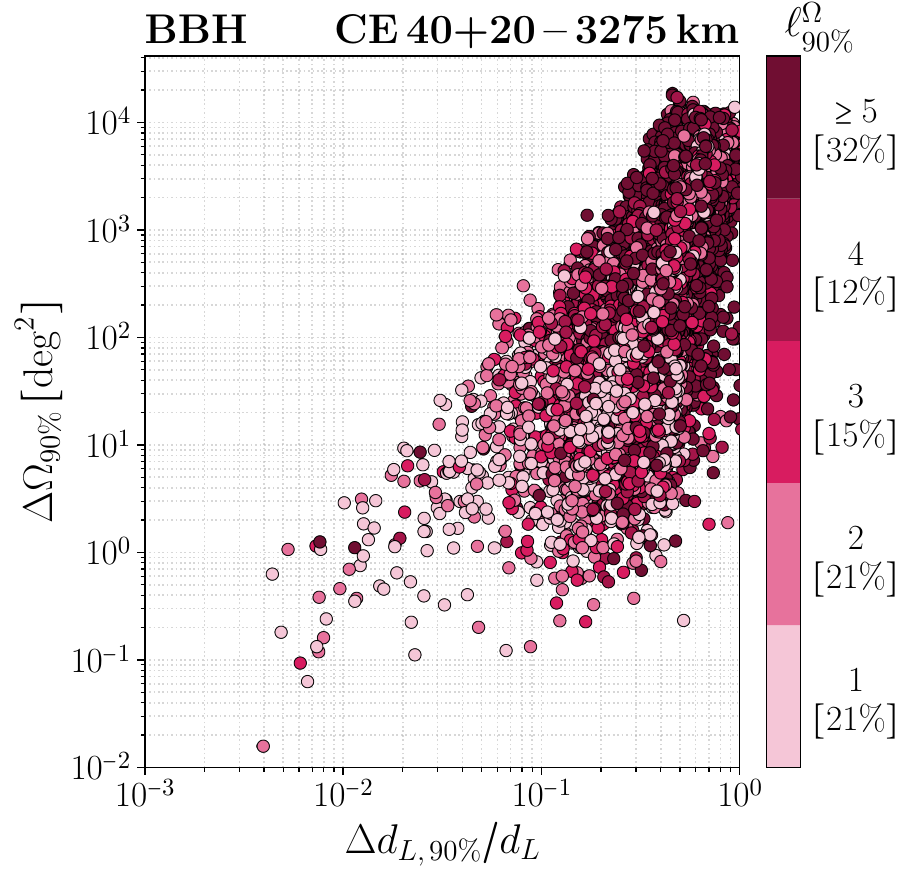} & \includegraphics[width=.32\linewidth]{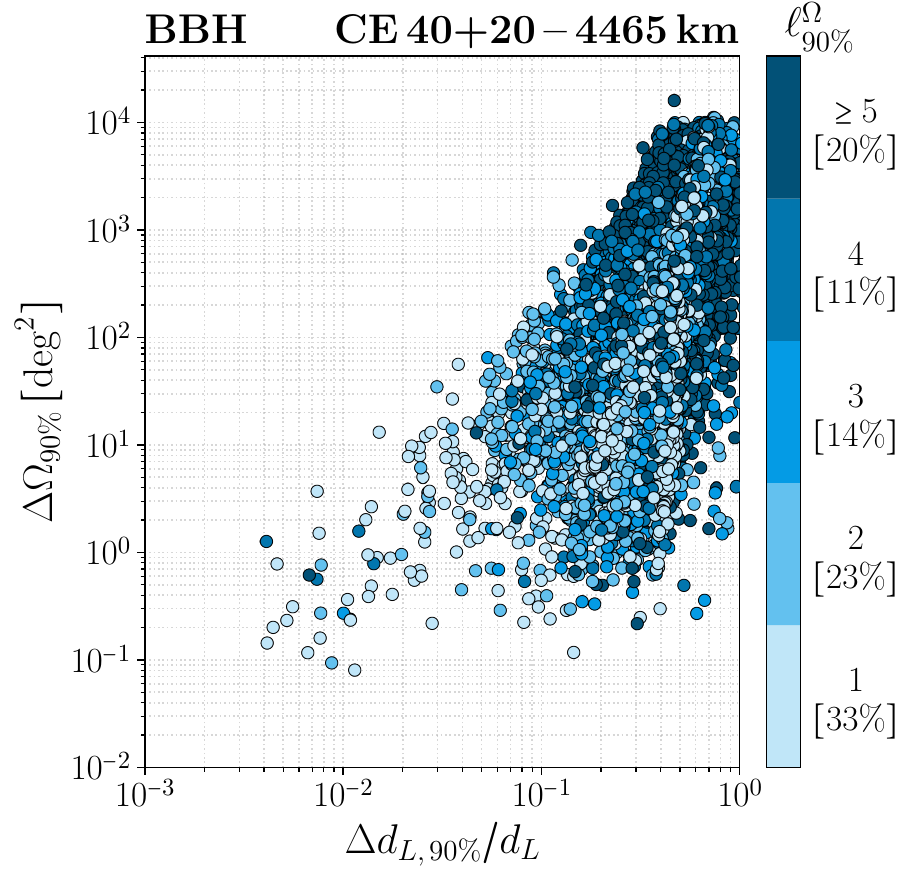}
    \end{tabular}
    \caption{Scatter plots of the sky localization and relative statistical uncertainty attainable on the luminosity distance at 90\% c.l. for a realistic BBH population observed with a network of 2 CE detectors. Each panel corresponds to a different detector baseline, while the points are colored according to the number of distinct modes in the sky localization of each source. In the colorbars, we report the percentage of sources with a given number of modes in the sky localization posterior.}
    \label{fig:scatter_OmegadL_2CE}
\end{figure*}

So far, we have focused on single sources with fixed detector-frame masses and SNRs. We now analyze the impact of the baseline on the localization of a realistic BBH population corresponding to 1\,yr of observations, consistent with the latest LVK analysis~\cite{LIGOScientific:2025pvj}. Let us stress that these results inherently depend on our specific choice of the astrophysical distribution of masses, spins, and redshifts, yet they provide a complementary view to the results discussed in previous sections, and serve as a meaningful benchmark to compare different detector configurations. 

In Fig.~\ref{fig:scatter_OmegadL_2CE} we show the distribution of 90\% sky localization and of the relative uncertainty on the luminosity distance for various CE baselines. The points are colored according to the number of modes in the sky localization posterior. The number of detected systems in our catalog with the chosen SNR cut is approximately $\sim\!1.03\times10^5$ events in a year for all the CE baselines. This corresponds to a fraction of $\sim\!88\%$ of the full catalog we simulate after excluding events with inspirals longer than 5\,min in the detector's band. 
The longest baseline results in more precise localization for the bulk of the observed systems and yields a large fraction of posteriors with unimodal structure. As the detectors get closer and closer, the number of events with multimodal localization posteriors increases dramatically, resulting in larger localization areas and a smaller fraction of events with narrow $d_L$ posteriors. The very best localized events are less impacted by the detectors' baseline. 

\begin{figure*}[tb]
    \centering
    \includegraphics[width=\linewidth]{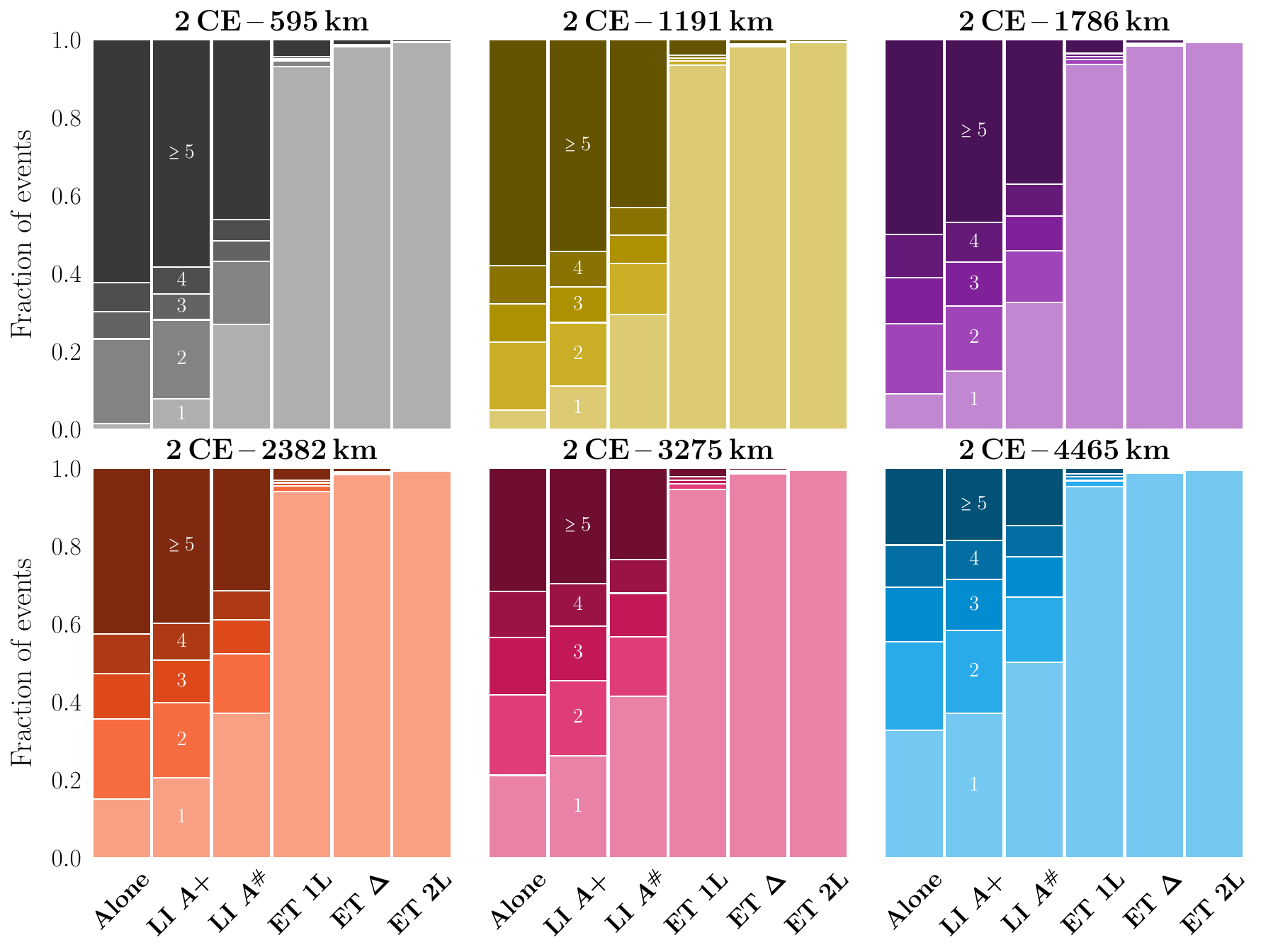}
    \caption{Fraction of events with a given number of modes in the sky localization posterior for combinations of two CE interferometers with different baselines, with various other detectors. From left to right, each bar corresponds to either a network of two CE detectors alone, or two CE with the addition of LIGO-India at $A+$ or $A^\#$ sensitivity, a single L-shaped ET observatory, a single ET in its triangular configuration, or two L-shaped ET interferometers. Each panel corresponds to a different baseline between the two CEs, while the color shadings from lightest to darkest refer to events with localization posteriors with 1 mode (lighter colors) and up to 5 or more modes (darker colors).
    }
    \label{fig:bar_modes_2CE_all_population}
\end{figure*}

\begin{table}[tb]
    \centering
    {\setlength{\tabcolsep}{3pt}
    \begin{tabularx}{\linewidth}{c|cccc|c}
    \toprule\midrule
    \multirow{3}{*}{\begin{tabular}{c}
        \textbf{2 CE}\\
        \textbf{baseline}
    \end{tabular}} & \multicolumn{4}{c|}{number of detections} & \multirow{3}{*}{\begin{tabular}{c}
        silver\\
        dark\\
        sirens
    \end{tabular}}\\
    \cmidrule{2-5}
    & in ZTF & in LSST & obs. by one & with & \\
    & FoV & FoV & hemisphere & $\ell^{d_L}>1$ & \\
    \midrule
    595\,km & 2953 & 421 & 9251 & 9317 & 28\\
    1191\,km & 6806 & 1138 & 22639 & 12698 & 43\\
    1786\,km & 10814 & 1856 & 37061 & 14231 & 48\\
    2382\,km & 14517 & 2556 & 50166 & 14015 & 51\\
    3275\,km & 20107 & 4149 & 60936 & 12418 & 69\\
    4465\,km & 26662 & 6849 & 74935 & 10671 & 106\\
    \midrule\bottomrule
    \end{tabularx}
    }
    \caption{For the different baselines of 2 CE instruments considered, we report: the number of detections whose localization area can be covered by a single pointing of ZTF and LSST, the number of detections with sky localization such that it could be entirely observable by a single instrument in one hemisphere, and the number of detections with a multimodal luminosity distance posterior. In the final column, we further report the number of silver dark sirens, i.e., detections with $\Delta\Omega_{90\%}\leq1\,{\rm deg}^2$.}
    \label{tab:2ce_pop_res}
\end{table}

As a meaningful metric, in Table~\ref{tab:2ce_pop_res} we again report the fraction of events with 90\% localization area entirely contained in the FoV of ZTF and LSST (approximated as a square in both cases), as well as the fraction of events with the modes of the sky localization posterior falling in a region of the sky that could be possible to observe with a single instrument in one hemisphere. For the longest baseline, about one quarter of all the detected events could be followed up optically with one pointing of ZTF, while three quarters lie in a single hemisphere and could thus be followed up by a single instrument. As the baseline decreases, fewer and fewer events can be followed up with a single pointing of ZTF, namely one out of seven events for the 2382\,km baseline and only $\sim\!3\%$ for the shortest one. A similar trend is observed in the fraction of systems with sky localization that could be covered by an instrument in a single hemisphere. Limiting ourselves again to events with $z\leq0.5$ observable with a single pointing of ZTF, we have that the longest baseline is able to deliver a number of 447 detections, lowering to 344 for the intermediate baseline of 2382\,km, and 228 for the shortest baseline. Finally, the events that could be followed up with a single pointing of LSST are instead less numerous: $\sim\!6.7\times10^3$ systems for the longest baseline, $\sim\!2.6\times10^3$ for the 2382\,km one, and only $\sim\!400$ for the shortest one.  

In Table~\ref{tab:2ce_pop_res}, we further report the number of detections exhibiting a multimodal luminosity distance posterior. In this case, the trend is not monotonic. The longest baseline produces $\sim\!10^4$ events with a multimodal $d_L$ structure. This number rises to $\sim\!1.4\times10^4$ for the 2382\,km baseline, and decreases again to $\sim\!9\times10^3$ for the detectors in their closest configuration. This behavior can be understood by considering two competing effects. The multimodalities in the marginal $d_L$ posterior arise due to multimodalities in the sky localization~\cite{Santoliquido:2025aiq}. The number of events with good localization lowers as we decrease the baseline, which explains the increase in the number of systems with multimodal distance posterior when going from a baseline of 4465\,km to 2382\,km. 
Decreasing the baseline even further, the sky localization posteriors get more degenerate and less informative, leading in turn to broader luminosity distance posteriors in which subdominant modes merge. In Fig.~\ref{fig:scatter_OmegadL_2CE}, this can be appreciated by looking at the different density of points in the region with small relative error on $d_L$ for different CE baselines. This effectively erases the multimodality in the inferred $d_L$ for some of the events, at the price of a less precise posterior.

Finally, in the rightmost column of Table~\ref{tab:2ce_pop_res} we report the number of so-called ``silver dark sirens''~\cite{Dang:2025vqx}, i.e., events localized better than $1\,{\rm deg}^2$ in the sky. These events are very promising for cosmological applications, as there is a limited number of potential host galaxies in such a small sky area. The number of silver dark sirens detected with two CE observatories in one year is modest: about 100 for the longest baseline considered, and 28 for the shortest one. We note that the localization area of silver dark sirens is comparable to the expected FoV at 1.4\,GHz of the full mid-frequency telescope for the Square Kilometre Array~\cite{2019arXiv191212699B}.
The number of  ``golden dark sirens,'' i.e., systems with  $\Delta\Omega_{90\%}\leq0.1\,{\rm deg}^2$~\cite{Borhanian:2020vyr,Gupta:2022fwd,Chen:2025qsl,Dang:2025vqx}, is even lower, with only $\sim \!2$ events per year for the two shortest baselines and 4 for the others. Nevertheless, it is noteworthy that, even for the shortest baseline, a network of two CE detectors can deliver a number of gold and silver dark sirens larger than a three-detector network with $A^\#$ sensitivity~\cite{Dang:2025vqx}. The localization area of golden dark sirens is comparable to or smaller than the FoV of facilities such as the \textit{Swift} X-Ray Telescope (XRT)~\cite{SWIFT:2005ngz}, the Jansky Very Large Array (VLA)~\cite{2009IEEEP..97.1448P} at 1.4\,GHz, the next-generation VLA (ngVLA)~\cite{2018ASPC..517....3M} at 2.4\,GHz, and of the Nancy Grace Roman Space Telescope~\cite{2019arXiv190205569A}. Hence, these events would also enable deep single-pointing observations with small FoV telescopes in search for potential EM counterparts.
We stress that we have excluded from our analysis events that spend longer than 5\,min in the detector frequency band, for which the localization is expected to be narrower and unimodal thanks to the long inspiral and self-triangulation allowed by the Earth's motion during the observation time. In this sense, our results are a lower bound on the number of silver and golden dark sirens. 

We now repeat the analysis on our 1\,yr BBH population, adding additional detectors to our two-CE network. We make the same approximation of Sec.~\ref{subsec:extended_net}, meaning that an event observed by at least three detectors with $\rm{SNR}\ge4$ in each of them has a unimodal sky localization posterior.
We consider the same combinations of two CE detectors plus LIGO-India with $A+$ or $A^\#$ sensitivity, a single L-shaped ET interferometer, two L-shaped ET instruments, and a single ET with triangular geometry. In Fig.~\ref{fig:bar_modes_2CE_all_population}, we show the fraction of events with a given number of modes in the sky localization posterior for the different detector networks. In this case, the addition of LIGO-India with $A+$ sensitivity only brings a minor improvement over the results with the two CEs, since the fraction of events with unimodal sky localization increases by only $\sim\!6\%$. On the other hand, LIGO-India with $A^\#$ sensitivity can increase the fraction of detected events with unimodal localization by $\sim\!17\%-25\%$, depending on the CE baseline. Considering combinations of the two CEs and ET instead, we find that even a single L-shaped ET increases the fraction of events with unimodal posteriors to about $94\%$ of the detected events, with only a mild dependence on the CE baseline. A combination of 2 CEs with ET in one of its complete configurations currently under active consideration, namely 2 L-shaped detectors or a single triangle, would instead result in almost all of the detected events featuring a unimodal sky localization, with the 2L configuration performing slightly better than the triangular one. 

\section{Conclusions}\label{sec:conclusions}

We have analyzed the impact of the baseline on the localization capabilities of a GW detector network comprising two CE detectors in the US. By considering two detectors with relative distances feasible on the US mainland and corresponding to light travel times ranging from 2\,ms to 15\,ms, we have first analyzed nonspinning BBH sources with fixed detector-frame total mass and SNR uniformly distributed in the sky, and then we considered a realistic population of short-duration BBH signals compatible with the latest LVK results. The first analysis does not depend on astrophysical assumptions; the second analysis is more model-dependent, but it provides a complementary benchmark to compare different configurations motivated by current observations. Since we focus on the potentially multimodal structure of the sky localization posterior, we have employed \texttt{BAYESTAR}, which allowed us to analyze a large number of sources while keeping the computational cost under control.

As expected, a longer baseline can deliver better sky localization. For events with fixed detector-frame total mass $M_{\rm tot}^{\rm det}$ up to $\sim\!100\,{\rm M}_\odot$, most observations would result in at most bimodal localization posteriors, irrespective of the SNR. A good fraction of these observations could fall entirely in the FoV of a telescope such as ZTF, allowing an efficient electromagnetic follow-up. Intermediate baselines with a light travel time of 11\,ms (i.e., a distance of 3275\,km, comparable with the one of LIGO Hanford and LIGO Livingston) or 8\,ms (2382\,km) are a reasonable compromise. Even lower baselines result in a larger degradation of the localization capabilities, especially for high-SNR events. These findings remain valid for binaries with unequal masses and different inclinations. 

The viability of a baseline with a light travel time of 8\,ms is confirmed by our population analysis of short-duration BBH signals. In this case, we find a sensible relative change in sky localization when we lower the baseline from 4465\,km to 2382\,km, with further degradation for lower baselines. Such short baselines can compromise some aspects of the science case of CE that rely on source localization, such as GW cosmology using statistical host identification techniques.

We have also studied multimodalities in the luminosity distance posterior, following the findings of Ref.~\cite{Santoliquido:2025aiq}, in which the authors reported that a detector configuration comprising 2 ET detectors in Europe could result in multimodal luminosity distance posteriors due to correlations with a multimodal localization posterior. In this case, the chosen baseline has a milder effect for fixed-mass injections, and the results are more dependent on the SNR. For a realistic population of sources, the events exhibiting a multimodal $d_L$ posterior are $\lesssim14\%$ of the whole catalog. %

The addition of a third detector eliminates multimodal sky localization posteriors, even for events that have small SNR in the third detector. This consideration allows us to estimate the localization capabilities of networks, including either LIGO-India or ET, in addition to the 2 CEs. The addition of LIGO-India to the network (especially if it will operate at $A^\#$ sensitivity) would remove multimodalities for a large fraction of the binaries up to $M_{\rm tot}^{\rm det}\sim200\,{\rm M}_\odot$ in our fixed-mass analysis and for about 25\% of the systems in our population analysis, regardless of the baseline. The addition of LIGO-India is more significant for shorter baselines. The addition of an ET detector in Europe solves the multimodality issue for the vast majority of the sources. Let us stress that our analysis did not include a duty cycle for the detectors. In reality, coordinated observations between the different detectors are crucial~\cite{Borhanian:2025uni}.

Our results have some limitations. While computationally efficient, \texttt{BAYESTAR}  can result in broader posteriors than a full Bayesian parameter estimation, and it does not include higher-order harmonics in the signal; see Appendix~\ref{app:highermodes}. 
Moreover, we focused on short-duration BBH signals that spend less than $5\,$min in the detector's frequency band. For longer signals, the modulation of the signal due to the Earth's motion would result in an improved localization even for a single detector. Our results are thus conservative and limited to a specific subset of events. Another possible extension of our work concerns the effect of the alignment between the detectors in the two-detector networks, which also affects the reconstruction of the sky localization and distance. In conclusion, we stress that the goal of this work is not to propose any realistic site or configuration for CE, but only to provide information that may be useful to inform those choices. 

\let\oldaddcontentsline\addcontentsline%
\renewcommand{\addcontentsline}[3]{}%
\begin{acknowledgments}
We thank Filippo Santoliquido for interesting discussions and comments on the manuscript. 
F.I., L.R., and E.B. are supported by NSF Grants No.~AST-2307146, No.~PHY-2513337, No.~PHY-090003, and No.~PHY-20043, by NASA Grant No.~21-ATP21-0010, by John Templeton Foundation Grant No.~62840, by the Simons Foundation [MPS-SIP-00001698, E.B.], by the Simons Foundation International [SFI-MPS-BH-00012593-02], and by Italian Ministry of Foreign Affairs and International Cooperation Grant No.~PGR01167.
This work was carried out at the Advanced Research Computing at Hopkins (ARCH) core facility (\url{https://www.arch.jhu.edu/}), which is supported by the NSF Grant No. OAC-1920103. 
The work of F.I. is supported by a Miller Postdoctoral Fellowship. A.C. acknowledges support from the NSF via grant No. AST-2431072.
B.S.S. is supported by NSF grants AST-2307147, PHY-2308886 and PHY-2309064. 
\end{acknowledgments}

\appendix

\begin{figure*}[tb]
    \centering
    \includegraphics[width=0.988\linewidth]{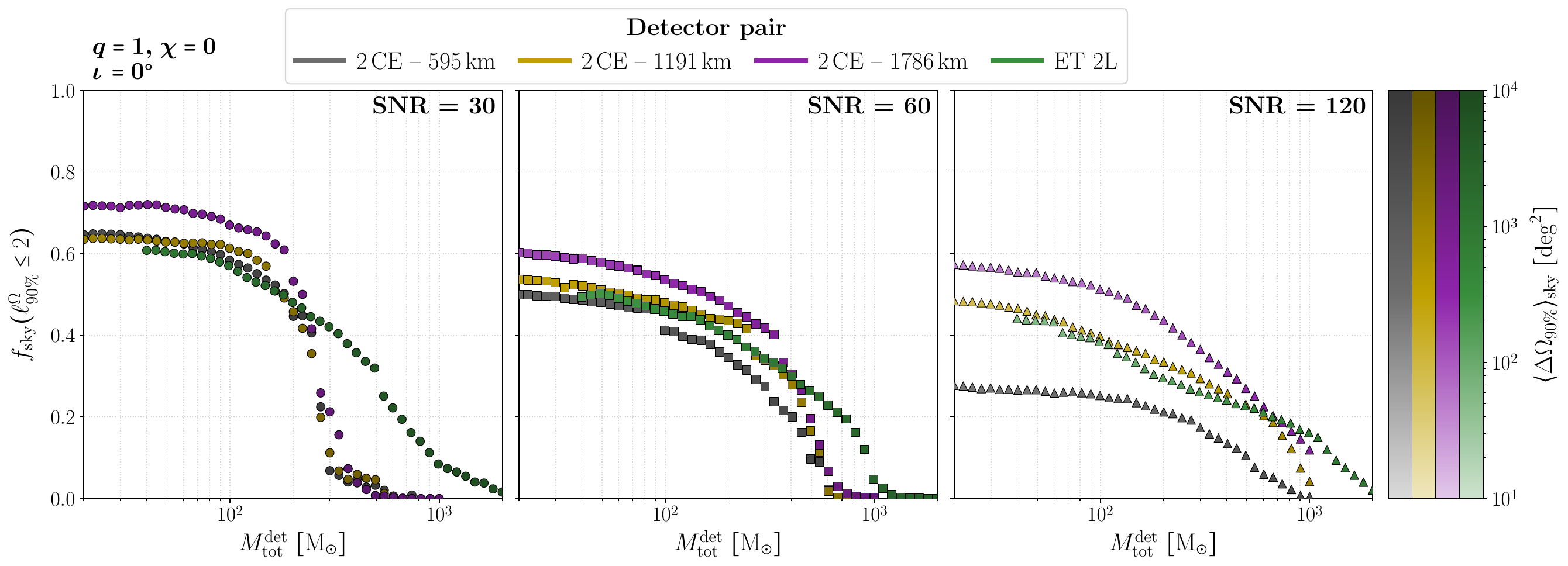}\\
    \includegraphics[width=0.988\linewidth]{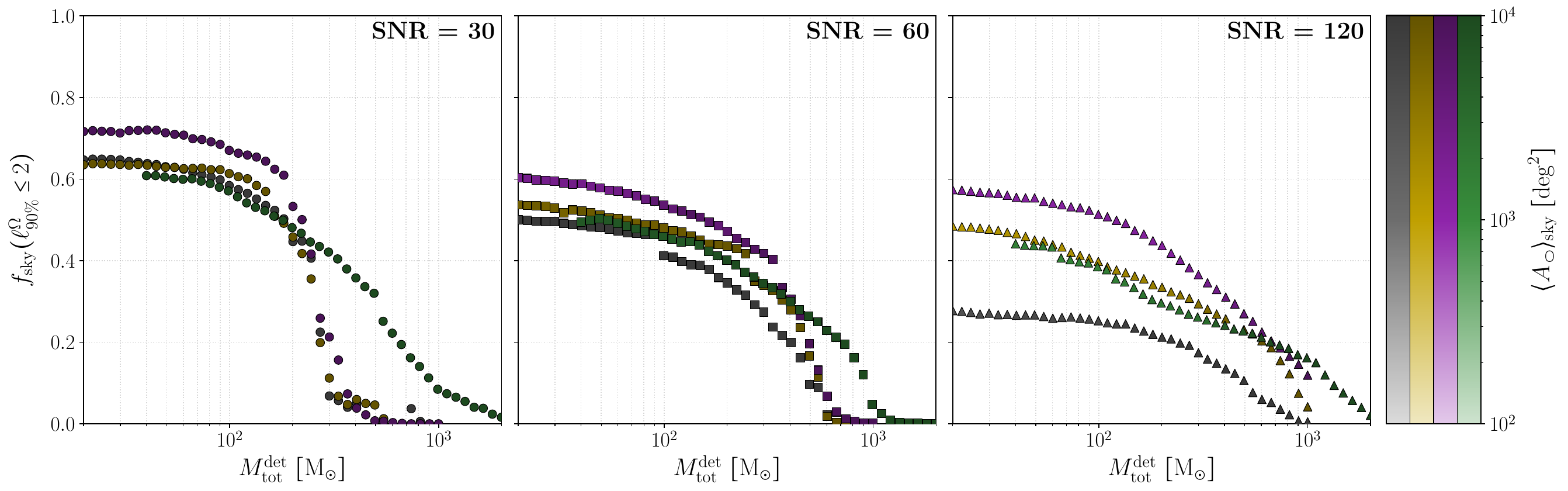} \\
    \includegraphics[width=\linewidth]{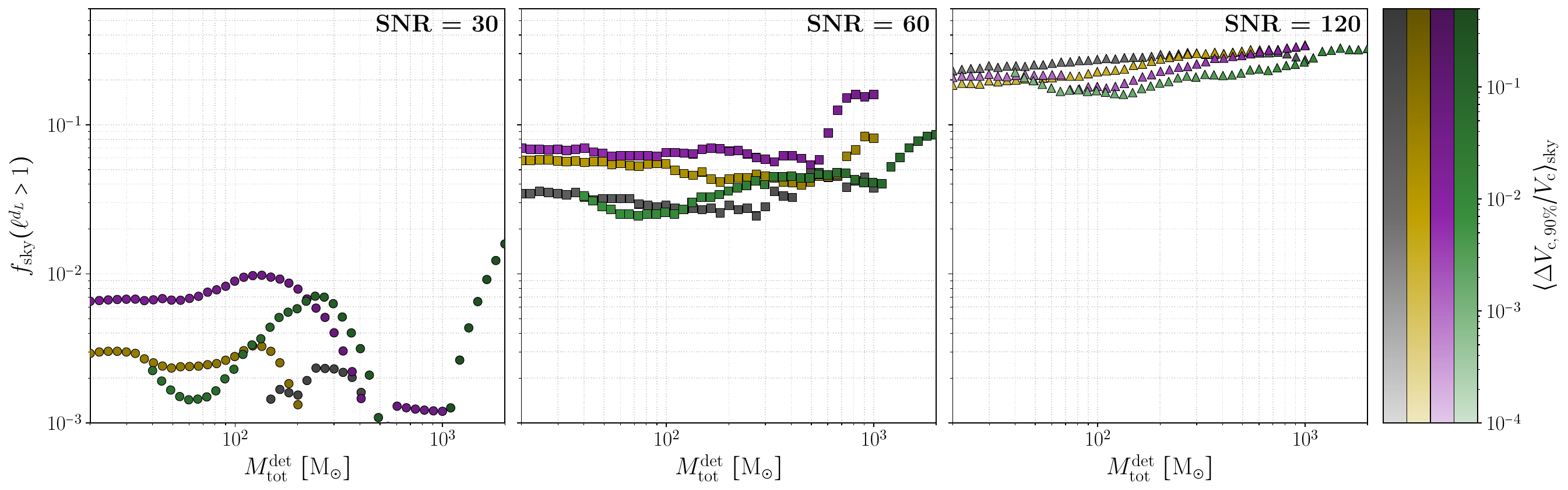}
    \caption{Same as in Fig.~\ref{fig:fraction_multimodalities_baseline} for the shorter baselines, and for a network of 2 L-shaped ET detectors.}
    \label{fig:fraction_multimodalities_et}
\end{figure*}

\section{The case of Einstein Telescope in the 2L configuration}\label{app:et}

\begin{table*}[tb]
    \centering
    {\setlength{\tabcolsep}{3.85pt}
    \begin{tabularx}{\linewidth}{c|c|ccc|ccc|ccc|ccc}
    \toprule\midrule
    \multirow{3}{*}{\textbf{detector}} & \multirow{3}{*}{SNR} &
    \multicolumn{9}{c|}{percentage of systems} & \multicolumn{3}{c}{\multirow{2}{*}{$\langle\theta_{{\rm sep},\,90\%}\rangle$}} \\
    \cmidrule{3-11}
    & & \multicolumn{3}{c|}{in ZTF FoV} & \multicolumn{3}{c|}{in LSST FoV} & \multicolumn{3}{c|}{obs. by one hemisphere} & \\
    & & $40\,{\rm M}_\odot$ & $200\,{\rm M}_\odot$ & $800\,{\rm M}_\odot$ & $40\,{\rm M}_\odot$ & $200\,{\rm M}_\odot$ & $800\,{\rm M}_\odot$ & $40\,{\rm M}_\odot$ & $200\,{\rm M}_\odot$ & $800\,{\rm M}_\odot$ & $40\,{\rm M}_\odot$ & $200\,{\rm M}_\odot$ & $800\,{\rm M}_\odot$ \\
    \midrule
    \multirow{3}{*}{ET 2L} & 30 & 5.5\% & 2.5\% & 0\% & 0\% & 0\% & 0\% & 64.4\% & 50.7\% & 4.9\% & 106$^\circ$ & 120$^\circ$ & 152$^\circ$ \\
    & 60 & 23.8\% & 12.7\% & 4.0\% & 5.1\% & 0.9\% & 0.0\% & 75.9\% & 70.4\% & 28.0\% & 87$^\circ$ & 98$^\circ$ & 134$^\circ$ \\
    & 120 & 48.5\% & 34.7\% & 12.7\% & 14.7\% & 11.6\% & 2.9\% & 82.7\% & 80.3\% & 60.5\% & 76$^\circ$ & 82$^\circ$ & 112$^\circ$ \\
    \midrule\bottomrule
    \end{tabularx}
    }
    \caption{Same as in Table~\ref{tab:2ce_grid_res}, but for a network of 2 L-shaped ET detectors. Here we consider binaries with higher detector-frame total masses of $M_{\rm tot}^{\rm det}\simeq40,\,200,\,800\,{\rm M}_\odot$.}
    \label{tab:2et_grid_res}
\end{table*}

In Sec.~\ref{subsec:baseline_var}, we analyzed the capabilities of a network of two CE detectors with different baselines. In this appendix, we repeat the same study for a network of 2 L-shaped misaligned ET detectors, one placed in Sardinia and one in the Meuse-Rhine Euroregion~\cite{Branchesi:2023mws}. We expect that placing one of the detectors in Sardinia and the second in the Lusatia candidate site would yield similar results, given the comparable baseline. We once again consider equal-mass, nonspinning binaries on a grid of 40 detector-frame masses with three representative SNR values. For each mass, we distribute the events uniformly on a $40\times40$ grid over the sky. Given the larger bandwidth of ET towards lower frequencies and the lower frequency cutoff of 3\,Hz we adopt, we extend the mass range towards higher masses, $M_{\rm tot}\in[40,\,2000]\,{\rm M}_\odot$, excluding events that spend more than $\tau(3\,{\rm Hz})\simeq5\,{\rm min}$ in band. In Fig.~\ref{fig:fraction_multimodalities_et}, we report our results for ET and compare them explicitly with the results for the CE baselines of 595\,km, 1191\,km, and 1786\,km shown in Fig.~\ref{fig:fraction_multimodalities_baseline} of the main text. We note how the distance of 1191\,km between CE detectors is comparable to the ET baseline.

The sky localization accuracy of a network with 2 ET detectors is qualitatively similar to two CEs with a 1191\,km baseline, albeit shifted towards higher masses, because the noise curve of ET is optimized for lower frequencies; see Fig.~\ref{fig:all_asds}. Results for the average separation between different modes, the fraction of injections falling in the FoV of ZTF and LSST, and the fraction of injections with localization that could be covered by an instrument in a single hemisphere are listed in Table~\ref{tab:2et_grid_res} for some representative mass values (twice the masses listed in Table~\ref{tab:2ce_grid_res}, because we are considering a different grid in masses for ET). Typical results are somewhere in between what we found for 2 CEs separated by a baseline of 1191\,km and 2 CEs separated by a baseline of 1786\,km, but shifted to higher mass. The average values across all masses and sky positions also have similar trends: 2.1\% (10.9\%, 28.0\%) of the injections could be observed with a single pointing of ZTF, and 86\% (71\%, 54\%) of them would be localized to different hemispheres for ${\rm SNR} = 30$, (60, 120), respectively. This is because ET has better sensitivity at low frequencies, thus systems can accumulate more SNR over the longer inspiral, leading to narrower localization areas. In this sense, the detector noise ASD is effectively ``compensating'' for the shorter baseline.

\begin{figure*}[tb]
    \includegraphics[width=.98\linewidth]{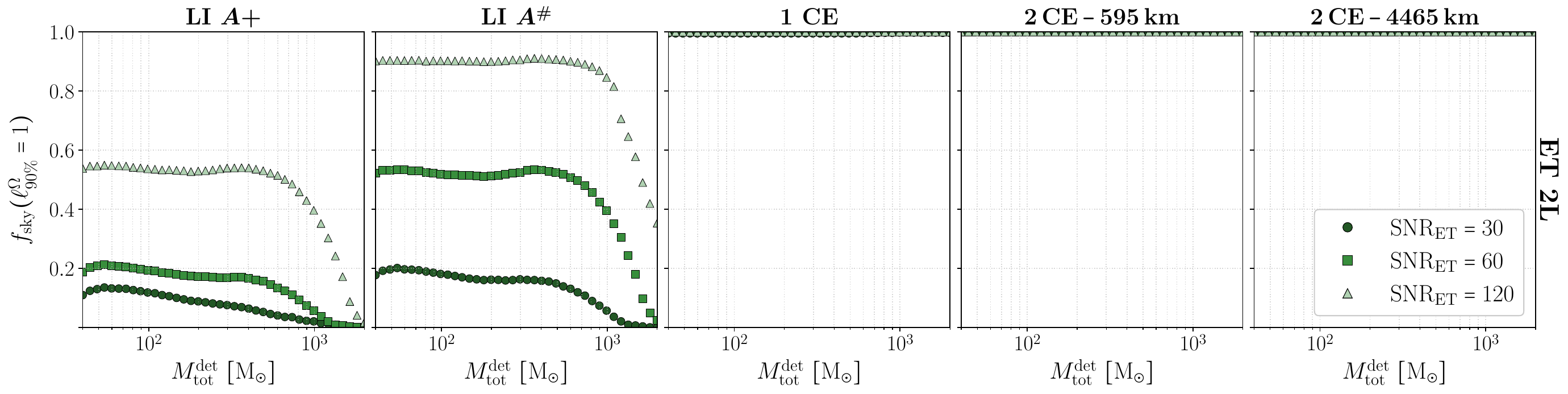} 
    \caption{Fraction of the points in the sky exhibiting a unimodal sky localization posterior as a function of the detector frame mass for different combinations of 2 L-shaped ET interferometers and other detectors (as indicated in each panel). We report the results for three SNRs in the 2 ET detectors: ${\rm SNR}=30$ (dark colored circles), ${\rm SNR}=60$ (medium colored squares), and ${\rm SNR}=120$ (light colored triangles). We consider equal-mass nonspinning binaries with face-on orientation.}
    \label{fig:fraction_unimodal_sky_deltaOmega_networks_ET}
\end{figure*}

\begin{figure}[tb]
    \centering
    \includegraphics[width=\linewidth]{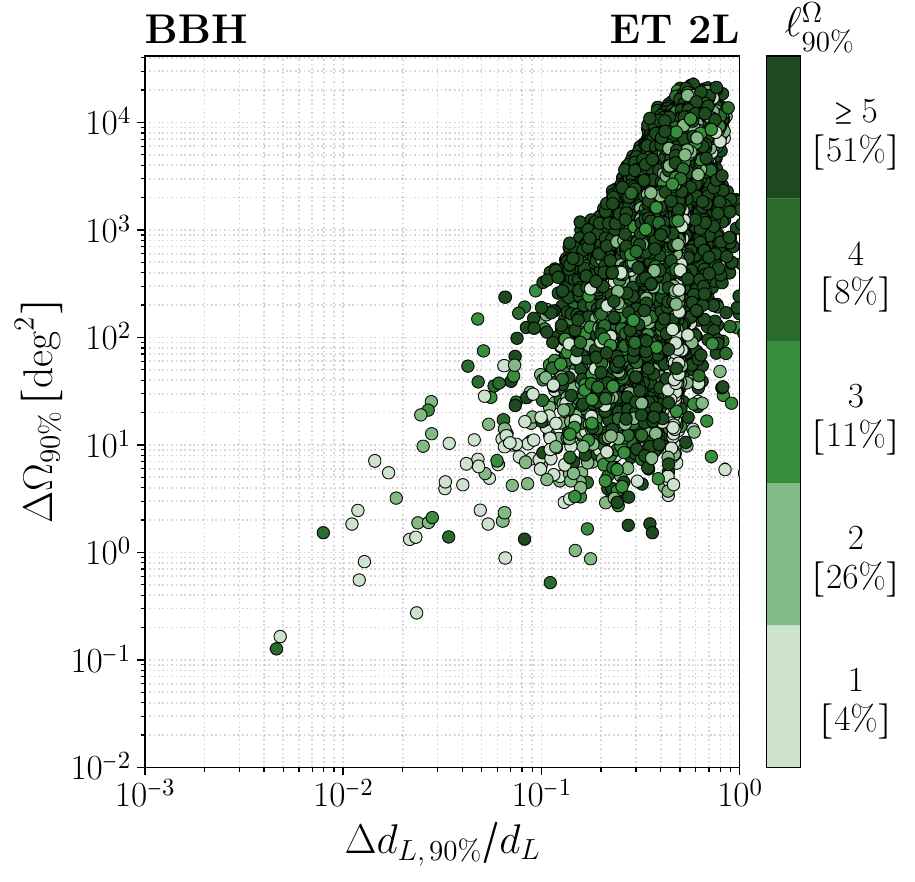}
    \caption{Same as Fig.~\ref{fig:scatter_OmegadL_2CE}, but for a network of 2 L-shaped ET detectors.}
    \label{fig:scatter_OmegadL_ET}
\end{figure}

\begin{figure}[tb]
    \centering
    \includegraphics[width=.67\linewidth]{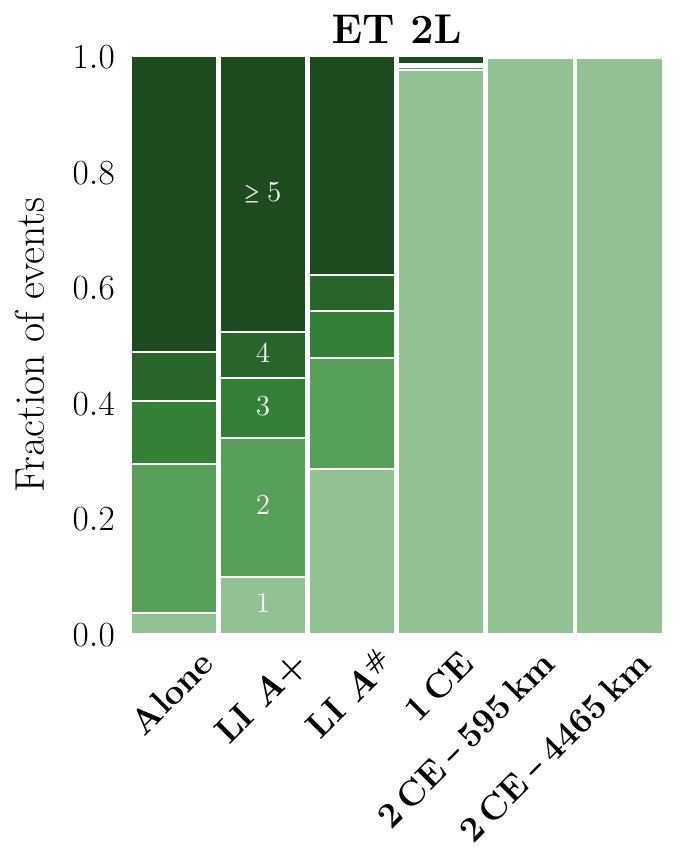}
    \caption{Same as in Fig.~\ref{fig:bar_modes_2CE_all_population} for a network of 2 L-shaped ET detectors. From left to right, each bar corresponds to either a network of two ET detectors alone, or two ET with the addition of either LIGO-India at $A+$ or $A^\#$ sensitivity, a single CE observatory with 40\,km arms, and two CE detectors with the shortest and longest baseline considered in the main text, respectively.}
    \label{fig:bar_modes_ET_population}
\end{figure}

Concerning the reconstruction of the distance, the 2L-configuration of ET results in less multimodal and narrower posteriors compared to two CE detectors. Once again, this can be explained by the different noise ASDs of the two detectors. ET is more sensitive at lower frequencies, while CE has higher sensitivity in the bucket $f\gtrsim10\,{\rm Hz}$ (see Fig.~\ref{fig:all_asds}). Therefore, for a signal with a given detector-frame total mass, ET accumulates more SNR at lower frequencies and observes the signal for a longer time, allowing it to better disentangle the two polarizations and thus improving the distance estimation. 

Adapting the procedure of Sec.~\ref{subsec:extended_net} to the two-detector ET configuration, we can compute the fraction of multimodal injections when we embed the two ETs in a global network including LIGO-India and CE. We consider combinations of ET with (\emph{i}) a single 40\,km CE placed in the northernmost position considered in the main text, and (\emph{ii}) two CE detectors, one of 40\,km and one of 20\,km arms, with the longest and shortest baselines considered in the main text. The results are reported in Fig.~\ref{fig:fraction_unimodal_sky_deltaOmega_networks_ET}. Similarly to the case of 2 CEs, the inclusion of LIGO-India already alleviates the multimodality issue for a large fraction of the binaries up to high total masses, with a steeper decline at high $M_{\rm tot}^{\rm det}$ because the mass grid now extends to higher masses. When we consider a global XG network, the inclusion of a single CE instrument is enough to remove multimodalities for practically all of the events.

We also repeat the population analysis of Sec.~\ref{subsec:loc_method_injs} for the two-detector ET configuration. We now have a smaller number of events compared to Sec.~\ref{subsec:pop_analysis}, since we still limit the observational time to 5\,min, and the low-frequency cutoff is set to 3\,Hz. The total number of events satisfying our threshold of 10 for the network SNR and 4 for the SNR in each detector is $\sim\!8.1\times10^4$, corresponding to a fraction of 81\% of the events in the full catalog after excluding systems with inspiral longer than 5\,min. Our results for the distribution of 90\% sky localization and relative uncertainty on the luminosity distance are reported in Fig.~\ref{fig:scatter_OmegadL_ET}. For this two-detector ET configuration, the number of events with localization falling entirely in the FoV of ZTF (LSST) is 4198 (573); 44954 events are localized within a region that could be covered by an instrument in a single hemisphere; and 6746 events have multimodal luminosity distance posteriors. This is somewhere in between the two-detector CE configurations with baselines of 1191\,km and 1786\,km for the localization, and again improved for the distance. In particular, a larger fraction of systems has a localization area with two distinct modes (rather than 3 or more) for ET. This can again be traced to the improved low-frequency sensitivity. We also find 8 silver dark sirens in the analysis. 

One should be cautious when comparing these numbers to those for configurations involving 2 CEs. We are explicitly excluding from the ET analysis a larger number of events with long inspirals, which would have good sky localization. Moreover, compared to Ref.~\cite{Santoliquido:2025aiq}, we find a lower fraction of events exhibiting a multimodal luminosity distance posterior and a multimodal sky localization posterior. This can be traced to two main factors: (\emph{i}) in our \texttt{BAYESTAR} analysis we include only the dominant emission mode of the signal, while in Ref.~\cite{Santoliquido:2025aiq} the authors employ \href{https://github.com/dingo-gw/dingo}{\texttt{DINGO}}, which exploits neural posterior estimation to estimate the parameters~\cite{Green:2020hst,Dax:2021tsq} and a waveform model that includes sub-dominant harmonics in the signal; (\emph{ii}) the authors of Ref.~\cite{Santoliquido:2025aiq} focus on a population of massive high-$z$ BBH mergers, for which the localization is particularly difficult, and consider $10^3$ events; in contrast, in our catalog we select the short-duration signals from a population compatible with the latest LVK results, thus probing a different mass range and redshift distribution, and we use $\sim\!100$ times more simulated injections.   

Finally, in Fig.~\ref{fig:bar_modes_ET_population}, we report the fraction of events with multimodal sky localization posteriors in the full 1-yr population when we embed the 2 ET detectors in a global network with LIGO-India and CE, adopting the same approximation as in Sec.~\ref{subsec:loc_method_injs}. The inclusion of LIGO-India with $A+$ ($A^\#$) sensitivity results in 7\% (25\%) more injections featuring a unimodal localization posterior. For a global network with 1 CE, only $\sim\!2.5\%$ of the injections have multimodal localization. For 2 ETs + 2 CEs, essentially all of the injections have unimodal localization, regardless of the CE baseline. Recall that we do not include a duty cycle for the detectors, but consider a continuous observing time of 1\,yr. We find some mild discrepancy with Ref.~\cite{Santoliquido:2025aiq} for the fraction of multimodal events, which can again be traced to differences in the populations and analysis methods that we adopted in our work.

\section{Including higher-order modes in the analysis}\label{app:highermodes}

As we stress in the main text, one of the limitations of our analysis concerns the modeling of higher-order (subdominant) harmonics in the signal, since only the dominant $(\ell=2, |m|=2)$ harmonic is used in \texttt{BAYESTAR}. Higher-order modes can help in breaking the degeneracy between distance and inclination in the signal for unequal mass $\iota\neq 0$ systems, leading to narrower and potentially less multimodal posteriors. To assess the impact of higher-order modes on our analysis, we randomly select a subsample of $10^3$ detectable events from the realistic population simulated in Sec.~\ref{subsec:pop_analysis}. We then re-evaluate the sky localization and luminosity distance posteriors for each event using a version of the \href{https://github.com/jroulet/cogwheel}{\texttt{cogwheel}} library~\cite{Roulet:2022kot,Islam:2022afg,Roulet:2024hwz} adapted for the \texttt{IAS-HM} search pipeline~\cite{Wadekar:2024zdq}, which includes the higher-order mode contribution $(\ell=3, |m|=3)$ and $(\ell=4, |m|=4)$ alongside the dominant harmonic $(\ell=2, |m|=2)$.

There are three inputs to this algorithm: (\emph{i}) $\langle d|h_{\ell,m}\rangle$, which corresponds to the input SNR timeseries for each of the mode templates; (\emph{ii}) $\langle h_{\ell,m} | h_{\ell',m'}\rangle$ which is the covariance matrix of the modes; (\emph{iii}) prior samples for mode SNR ratios according to an assumed astrophysical distribution 
\begin{equation}\label{eq:Rlm}
    R_{\ell m} \equiv \frac{\langle h_{\ell m}(f)|h_{\ell m}(f)\rangle^{1/2}}{\langle h_{22}(f)|h_{22}(f)\rangle^{1/2}}\,.
\end{equation}
See Sec.~3A of Ref.~\cite{Wadekar:2024zdq} for further details on this algorithm. In the previous expressions $\langle \cdot| \cdot \rangle$ denotes the noise weighted inner product, defined for two generic frequency domain signals $a(f)$ and $b(f)$ as
\begin{equation}
    \langle a(f) | b(f) \rangle = 4 {\rm Re} \int \dfrac{a^\ast(f) b(f)}{S_n(f)}\ {\rm d}f \,, 
\end{equation}
where $S_n(f)$ denotes the detector noise power spectral density (PSD).
\texttt{cogwheel} is used to efficiently marginalize over the extrinsic binary parameters and the subset of intrinsic parameters which affect the mode amplitude ratios in Eq.~\eqref{eq:Rlm}.
The integral over different extrinsic parameters is computed as follows: over the luminosity distance by interpolating a precomputed table; over the reference $\phi_0$ by trapezoid quadrature; over the remaining extrinsic parameters inclination, sky location, and polarization using adaptive importance sampling. We refer the reader to Ref.~\cite{Roulet:2024hwz} for further details on the marginalization procedure.

\begin{figure}[t]
    \centering
    \includegraphics[width=\linewidth]{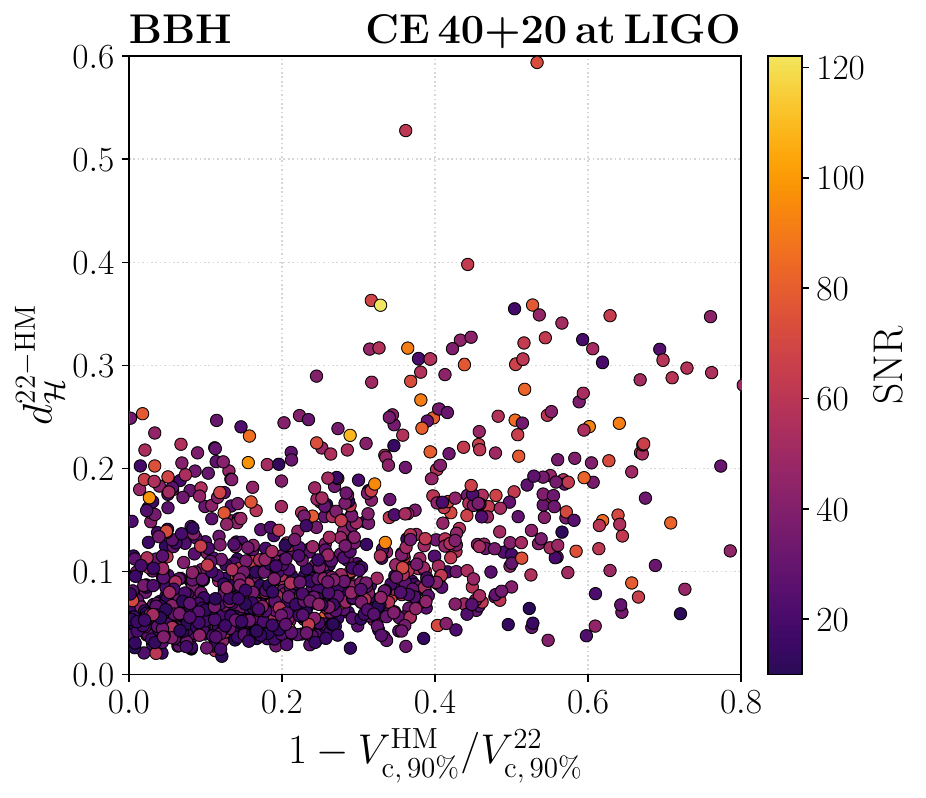}
    \caption{Scatter plot of the relative difference in localization comoving volume and Hellinger distance between the results obtained modeling only the dominant $(2,\,2)$ harmonic in the GW signal, and modeling also the subdominant $(3,\,3)$ and $(4,\,4)$ harmonics. We report the results for a subset of $10^3$ BBH events observed at a network comprising 2 CE detectors with the position and orientation of the current LIGO interferometers. The color is proportional to the SNR of each event.}
    \label{fig:scatter_VcHellinger_CEatLIGO_allevents}
\end{figure}

Once the marginalization integral is computed, the posterior samples are straightforward to obtain by using importance resampling over the quasi-Monte Carlo samples used for the marginalization integral. Overall, a multi-detector trigger needs $\sim 0.2$ sec for calculating the full marginalization integral and thus is efficient for low-latency analysis \cite{Iac26_inprep} and for use over large population catalogs such as the ones used in this paper.

For ease of implementation, we conduct this study using the same CE noise curves employed in the main text, but we place the detectors at the location and orientation of the LIGO Hanford (for the 40\,km detector) and LIGO Livingston (for the 20\,km detector) interferometers. We model the different mode waveforms using the \textsc{IMRPhenomXHM} waveform model~\cite{Garcia-Quiros:2020qpx}. When comparing with the findings in the rest of the paper, one should thus keep in mind that the results presented here are for nearly (anti)parallel detectors, for which disentangling the two polarizations of a GW signal is more difficult compared to the misaligned configuration employed in the main text (see e.g. Ref.~\cite{Branchesi:2023mws}).

In Fig.~\ref{fig:scatter_VcHellinger_CEatLIGO_allevents}, on the $x$-axis we show the ratio of the 90\% localization comoving volume obtained with and without the inclusion of higher-order harmonics for the chosen event. On the $y$-axis, we plot the Hellinger distance between the three-dimensional posteriors in right ascension, declination, and luminosity distance~\cite{Hellinger1909}
\begin{equation}
    d_{\cal H}^{\rm 22-HM} = d_{\cal H} (p^{22},\, p^{\rm HM}) = \sqrt{1 -\!\! \int \sqrt{p^{22} \; p^{\rm HM}} \,{\rm d} V}\,,\!
\end{equation}
where $p^{22}$ and $p^{\rm HM}$ denote the posterior distributions with and without the inclusion of higher-order harmonics. This quantity encodes information about the similarity of two probability distributions: two identical distributions give $d_{\cal H}=0$, while two completely different distributions result in $d_{\cal H}=1$. For the majority of the events, we find that the improvement in the localization volume obtained including higher-order modes is modest ($\lesssim 20\%$), and that the 3D posteriors are relatively similar ($d_{\cal H}^{\rm 22-HM}\lesssim 0.1$). For a subset of high-SNR events with moderate inclinations, the inclusion of subdominant harmonics is instead more relevant, resulting in appreciably narrower distributions for the luminosity distance. However, even in these high-SNR cases, we find the impact on the sky localization posterior to be significantly more modest compared to the improvement in the reconstruction of $d_L$, except for a few events. 

Regarding multimodalities, we find $\sim\!67\%$ of the simulated events to feature a 90\% sky localization posterior with more than two distinct modes. This fraction is larger than the one found in the main text with a CE baseline of 3275\,km, comparable to the distance between the two LIGO detectors. This difference can be attributed to the change in orientations between the detectors considered here. More importantly, we find that the inclusion of higher-order harmonics in the analysis causes a reduction in the number of sky localization modes for only $\sim\!8\%$ of the cases.

In Fig.~\ref{fig:scatter_VcHellinger_CEatLIGO_allevents}, we also see some points with relatively large Hellinger distance but modest improvement in localization volume. These points correspond to cases in which the inclusion of subdominant harmonics can alleviate minor multimodalities in the $d_L$ posterior.
Indeed, for this configuration, we find a fraction of $\sim\!24\%$ of the events featuring a multimodal luminosity distance posterior. This can again be attributed to the relative orientation between the detectors used for this analysis. In this case, the inclusion of higher-order harmonics is able to alleviate (albeit not always completely remove) the multimodality for $\sim\!60\%$ of them, bringing the fraction of events with multimodal posteriors down to $\sim\!12\%$. This is in line with the findings of Ref.~\cite{Santoliquido:2025aiq}, in which the authors find multimodal luminosity distance posteriors despite the inclusion of subdominant harmonics in the analysis. This confirms that our population analysis should be considered pessimistic regarding the issue of multimodality in the luminosity distance posteriors. 

\bibliography{2det_loc}

\end{document}